\documentclass[journal=jpccck,manuscript=article]{achemso}

\usepackage{chemformula} 
\usepackage[T1]{fontenc} 
\SectionNumbersOn
\usepackage{siunitx}
\usepackage{bm}%
\usepackage{graphicx}
\usepackage[utf8]{inputenc}
\usepackage{epstopdf}

\author{Jan Weinreich}
\affiliation{Universit\"{a}t G\"{o}ttingen, Institut f\"{u}r Physikalische Chemie, Theoretische Chemie, Tammannstra\ss{}e 6, 37077 G\"{o}ttingen, Germany}
\author{Anton R\"omer}
\affiliation{Universit\"{a}t G\"{o}ttingen, Institut f\"{u}r Physikalische Chemie, Theoretische Chemie, Tammannstra\ss{}e 6, 37077 G\"{o}ttingen, Germany}
\author{Mart\'{i}n Leandro Paleico}
\affiliation{Universit\"{a}t G\"{o}ttingen, Institut f\"{u}r Physikalische Chemie, Theoretische Chemie, Tammannstra\ss{}e 6, 37077 G\"{o}ttingen, Germany}
\author{J\"org Behler}
\affiliation{Universit\"{a}t G\"{o}ttingen, Institut f\"{u}r Physikalische Chemie, Theoretische Chemie, Tammannstra\ss{}e 6, 37077 G\"{o}ttingen, Germany}
\altaffiliation{International Center for Advanced Studies of Energy Conversion (ICASEC), Universit\"at G\"ottingen, Tammannstra\ss{}e 6, 37077 G\"ottingen, Germany}
\email{joerg.behler@uni-goettingen.de}

\title{Properties of $\alpha$-Brass Nanoparticles I: Neural Network 
Potential Energy Surface}

\begin{document}

\begin{abstract}
Binary metal clusters are of high interest for applications in heterogeneous catalysis and have received much attention in recent years. To gain insights into their structure and composition at the atomic scale, computer simulations can provide valuable information if reliable interatomic potentials are available. 
In this paper we describe the construction of a high-dimensional neural network potential (HDNNP) intended for simulations of large brass nanoparticles with thousands of atoms, which is also applicable to bulk $\alpha$-brass and its surfaces. 
The HDNNP, which is based on reference data obtained from density-functional theory calculations, is very accurate with a root mean square error of $\SI{1.7}{\milli \electronvolt}$/atom for total energies and $\SI{39}{\milli \electronvolt \per \angstrom}$ for the forces of structures not included in the training set. The potential has been thoroughly validated for a wide range of energetic and structural properties of bulk $\alpha$-brass, its surfaces as well as clusters of different size and composition demonstrating its suitability for large-scale molecular dynamics and Monte Carlo simulations with first principles accuracy. 
\end{abstract}

\section{Introduction}\label{sec:introduction}

The interest in metal and alloy nanoparticles is constantly growing~\cite{P5542,P1842,P5544,P3458}. 
They often exhibit a high chemical activity and are characterized by distinct structural features like size, shape, and chemical composition, which can be locally very inhomogeneous e.g. in core-shell structures \cite{P5699}. New synthesis techniques now enable controlled fabrication at the nanometer-scale \cite{Xia2009,P5545, C4CS00362D}, resulting in an increasingly diverse use reaching from industrial catalysis \cite{Vines2014} to medical applications \cite{Mody2010}.

Brass, an alloy of copper and zinc, has become a textbook example of binary metallic compounds. 
As a result of extensive studies the complete temperature- and composition-dependent phase diagram of brass is well known \cite{brass,Martienssen2005}. In terms of composition it is framed by the two extreme cases of pure face-centered-cubic (fcc) copper and hexagonal close-packed (hcp) zinc. For zinc atom fractions lower than approximately $40 \%$ the most stable form of brass is the $\alpha$-form, a substitutional alloy possessing fcc structure \cite{brass}, which  will be the focus of this work. 
For increasing zinc concentrations a series of other phases, the $\beta$-, $\gamma$-, $\delta$-, $\varepsilon$- and $\eta$ phases of brass, are formed~\cite{brass}.
In addition to detailed experimental studies, several computer simulations for bulk brass have been reported \cite{p2119,osti_1222964,p5553}.

In spite of this thorough characterization of the rich phase diagram of bulk brass, the even more complex energy landscape of brass nanoparticles is still largely unexplored. Only a few theoretical studies of small brass clusters exist in the literature \cite{Botticelli2008,P5555} and very little is known about the structure and composition of larger brass nanoparticles. 
This is surprising as the CuZn system is very interesting for technological applications, not only because most other copper nano-alloys like CuPt\cite{P3458}, or CuAu \cite{Bracey2009} are much more expensive, but also because brass formation has been suggested to play an important role in large-scale industrial processes like methanol synthesis~\cite{P3221,P5546}.

An accurate representation of the potential energy surface (PES) is essential for identifying stable structures and for the characterization of their properties in computer simulations. Therefore, electronic structure calculations like density functional theory (DFT) are the method of choice for theoretical studies, and DFT has been successfully applied to investigate numerous small metal clusters \cite{Salazar-Villanueva2006,Botticelli2008,2012Nanos,doi:10.1021/jp982775o}.
However, many important tasks, like global structure optimization or the calculation of thermodynamic properties, require the energy evaluation of thousands to millions of configurations, which is not
feasible when applying computationally demanding methods like DFT directly. Consequently, simulations at the DFT level are restricted to very small clusters, while for systematic studies of larger systems, less-demanding atomistic potentials are needed, which provide a direct analytic functional relation between the atomic positions and the potential energy.

In recent decades many types of atomistic potentials have been developed, mainly for bulk materials but also for clusters, such as the embedded atom method \cite{P0342}, Gupta many-body potentials \cite{Qin}, Finnis-Sinclair potentials \cite{P2990} or the cluster expansion method \cite{P5557}. These potentials are fast to evaluate, but in some cases it has been demonstrated that empirical potentials relying on physical approximations can provide a qualitatively wrong topology of the PES, yielding spurious local minima for clusters~\cite{P5521}.

A rather recent approach to construct very efficient PESs with first principles accuracy relies on machine learning methods, and starting with the seminal work of Doren and coworkers in 1995~\cite{P0316} many different types of machine learning potentials (MLPs)~\cite{P4885,P4263} have been proposed, including neural networks~\cite{P0421,P1388,P2559,P5366,P3033,P4945,P5596, GuoNeuralPolynomial}, Gaussian approximation potentials (GAPs)~\cite{P2630}, moment tensor potentials~\cite{P4862}, spectral neighbor analysis potentials \cite{P4644} and Kernel-based methods \cite{felixFCHL,operatorsQM,P4896}. 
One of the most frequently used types of MLPs are high-dimensional neural network potentials (HDNNPs) proposed by Behler and Parrinello in 2007~\cite{P1174,P5128}, which have already been applied successfully to a number of systems related to the present work, from copper~\cite{P3114,P3251} and other metal and alloy clusters~\cite{P4470,P4474,P4028} via surfaces~\cite{P3114,P4974} to metal clusters supported at oxide surfaces~\cite{P3827,P4475}.

This paper is the first of a series of two papers. Here, we will focus on the generation and validation of a HDNNP for brass nanoparticles applicable to very large systems starting from about 75 atoms up to many thousands of atoms.
After a short summary of the underlying methodology in sec.~\ref{sec:methods} and of the construction of the PES in sec.~\ref{sec:comp_details} we present the obtained HDNNP and detailed tests for its validation in sec.~\ref{sec:results}. We have investigated a wide range of physical properties for clusters of varying size and composition, as well as bulk and surface structures. We conclude that the obtained HDNNP allows to perform large-scale simulations with close-to first principles accuracy at a small fraction of the computational costs of DFT calculations.
The results of simulations employing this potential will be presented in Ref.~\citenum{JanPaper2}.

\section{Methods}\label{sec:methods}

In this work we use high-dimensional neural network potentials as introduced by Behler and Parrinello in 2007\cite{P1174} to construct a PES for large brass nanoparticles containing several thousand atoms. The method itself has been described in great detail elsewhere \cite{P4444,P4106} and here we give only a concise summary. 

In the HDNNP method, the total energy $E_{\rm tot}$ of a system containing $N$ atoms is given by a sum over atomic energy contributions $E_{i}$,

\begin{align}
   E_{\rm tot} = \sum_{i=1}^{N} E_{i}~,
\end{align}

which depend on the local chemical environments of the atoms $i$ that are defined by a cutoff radius $R_{\rm c}$. To ensure that all relevant atomic interactions are included, the cutoff radius has to be sufficiently large, and typically values between 6 and 10 \AA{} are used, such that the $E_{i}$ are effectively functions of the positions of a large number of neighboring atoms. 

The atomic energy contributions $E_i$ are provided by atomic neural networks (NNs) as a function of many-body atom-centered symmetry functions (ACSFs)\cite{doi:10.1063/1.3553717}, which describe the geometric arrangements of the neighboring atoms within the cutoff spheres. There is one type of atomic neural network with fixed architecture (i.e. number of layers and neurons per layer) per element in the system to ensure the chemical equivalence of all atoms of the same element. Each element-specific atomic neural network is then replicated as many times as atoms of the respective element are present in the system, while the structural input, i.e. the numerical values of the ACSFs, and atomic energy output values of each of these replicas depend on the atomic positions. By design, HDNNPs fulfill all physically mandatory invariances of the PES, i.e. translational, rotational and permutational invariance, exactly.

The parameters of the atomic neural networks are determined iteratively by minimizing the errors of energies and forces in a reference data set of structures covering the configuration space relevant for the intended simulations. Often DFT is used as the reference method. We note that only total energies are required, while individual atomic energies, which are not physical observables, are not needed. Since the atomic energies and forces only depend on the local environments, a HDNNP can be trained using rather small periodic and/or non-periodic structures typically containing only up to 200 atoms, but it can then be applied to much larger systems. This enables large-scale simulations with the accuracy of the underlying reference electronic structure method at a fraction of the computational costs.

\section{Computational Details}\label{sec:comp_details}

\subsection{Density Functional Theory}

The reference DFT calculations have been performed with the Perdew-Burke-Ernzerhof\cite{Perdew1996a} (PBE) exchange-correlation functional employing the electronic structure code VASP-5.3 \cite{Kresse1999,P0156}.
The target accuracy of the total energy of a HDNNP with respect to DFT is a few meV per atom, which thus defines the required convergence criterion for the DFT calculations. Convergence tests with respect to the number of k-points showed that 
in order to fulfill this criterion a k-point grid of $12 \times 12 \times 12$ is needed for a conventional four-atom copper fcc unit cell with a lattice constant of about $\SI{3.63}{\angstrom}$ along with a plane wave cutoff energy of $\SI{500}{\electronvolt}$ and projector augmented wave potentials\cite{Blochl1994,Kresse1999}.
Larger systems have been calculated using an adapted k-point grid corresponding to the same k-point density. The $\Gamma$-point centered k-point grids have been constructed employing the Monkhorst-Pack scheme \cite{Monkhorst1976}. 

For surface calculations, 4-14 layer slabs with a total vacuum thickness of at least 8~\AA{} have been used. In the case of cluster calculations, which have also been treated in a periodic setup, the periodic images of the clusters have been separated by at least $\SI{8}{\angstrom}$ in all three spatial directions. The convergence of very large clusters with diameters of $d\approx\SI{22}{\angstrom}$, which have been used to include specific atomic environments in the data set, has been extensively tested and we found that using the $\Gamma$-point only is sufficient to reach the required convergence level.

\subsection{Description of the Atomic Environments} \label{sec:symmetry}

\begin{figure*}[htb]
\centering
\includegraphics[width=0.5\textwidth]{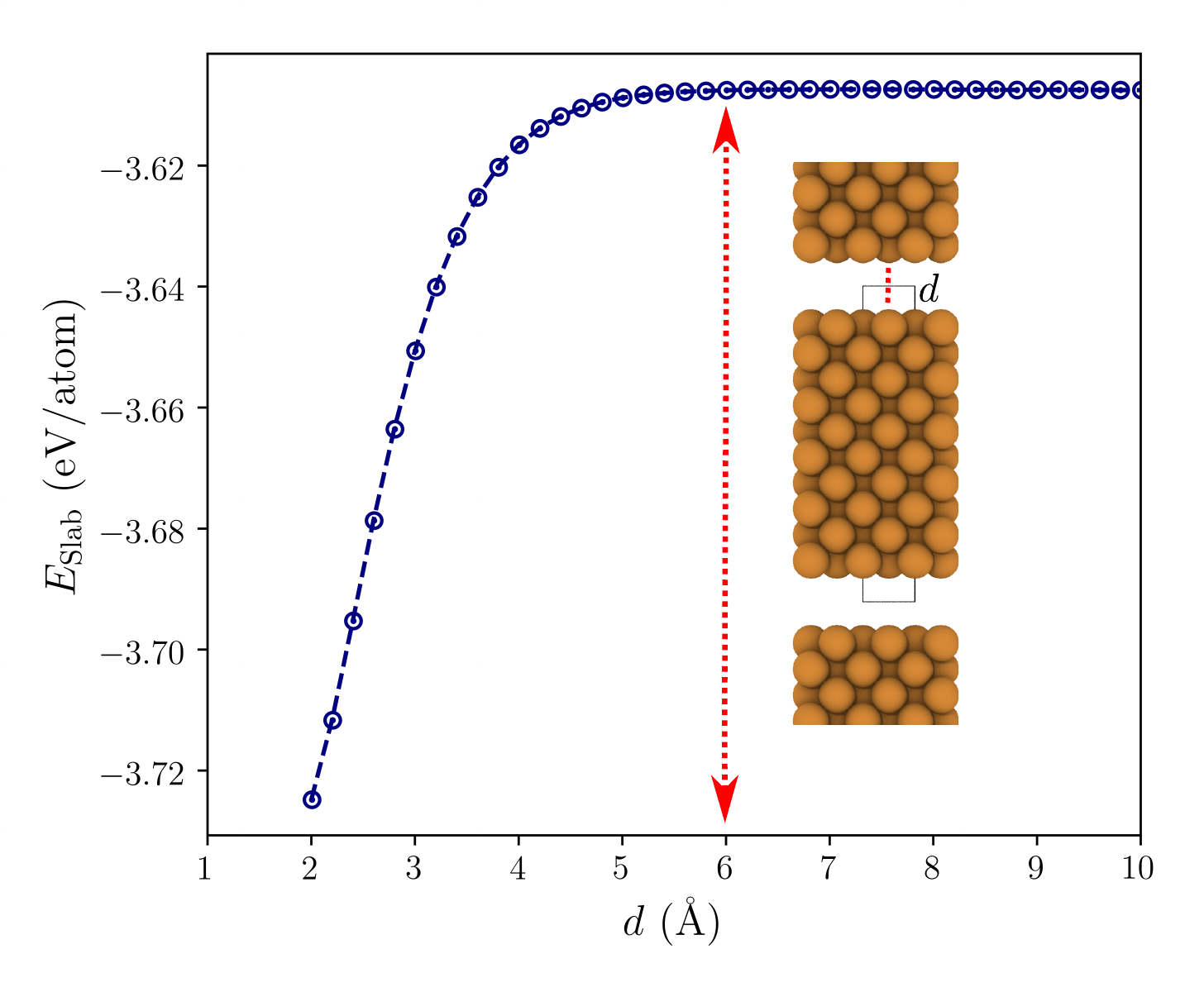}
\caption{DFT energy of a Cu(100) slab as a function of vacuum sizes $d$ which is the distance between the periodic images of the copper slab. The distance $d$ is also defined in the inset of this figure where a unit cell (black rectangle) of the slab and its two next periodic images in the positive and negative $\hat{z}$ direction are shown.}
    \label{fig:convergence}
\end{figure*}

In the present work we use two types of ACSFs~\cite{doi:10.1063/1.3553717} for describing the local atomic environments, radial symmetry functions (type 2) defined as
\begin{eqnarray}
 G_{i}^{2} (R_{ij}) = \sum_{j} e^{-\eta (R_{ij} - R_{s})^2} \cdot f_{\text{c}}(R_{ij}) ~,
 \label{eq:sym1}
\end{eqnarray}
where $R_{ij}$ is the distance of neighbor atom $j$ from the central atom $i$,
and angular functions (type 3),

 \begin{align}
 G_{i}^{3} & (R_{ij},R_{ik},R_{jk},\theta_{ijk})  =  2^{1-\zeta} \sum_{j} \sum_{k > j} (1+ \lambda \cos{(\theta_{ijk})})^{\zeta} \nonumber \\
 & \cdot f_{\text{c}} (R_{ij})  f_{\text{c}} (R_{ik})  f_{\text{c}} (R_{jk}) \cdot e^{-\eta (R_{ij}^2 + R_{ik}^2 + R_{jk}^2)} ~
 \label{eq:sym2}
\end{align}

where $\theta_{ijk}$ is the angle enclosed by the distances $R_{ij}$ and $R_{ik}$ to two neighboring atoms $j$ and $k$.
The cutoff function $ f_{\text{c}} (R_{ij})$ is defined as
\begin{eqnarray}
    f_{\text{c}} (R_{ij}) = 
\begin{cases}~
\frac{1}{2} \cdot
\left( 
\cos{
\left( \frac{\pi R_{ij}}{R_{c}
}
\right)} +1 \right) 
 & \text{for }  R_{ij}  \leq R_{\text{c}} \\
0   &  \text{for }  R_{ij}  	> R_{\text{c}} ~.
\end{cases}
\end{eqnarray}
The Gaussian exponents $\eta$ control the effective spatial range of the symmetry functions, $R_s$ can be used to shift the centers of the Gaussians in the radial functions, $\zeta$ is a parameter to control the angular resolution of the angular functions and $\lambda=\pm 1$ defines the positions of the extrema of the cosine functions in Eq.~\ref{eq:sym2}. The parameters of the symmetry functions used in the present work are listed in Table~\ref{tab:fit_paras}. This set of functions represents a subset of the ACSFs that have previously been used to construct a HDNNP for copper~\cite{P3114}. 
The cutoff radius of $\SI{6}{\angstrom}$ is large enough to include the interaction of a central atom with about 75 neighboring atoms in a bulk-like environment, while the number of neighbors of surface atoms is generally lower and can strongly vary depending on the specific atomic configuration. 
In order to validate that the chosen cutoff is sufficiently large we performed DFT convergence test calculations using the vacuum size $d$ of a Cu(100) slab. We found (s. Fig~\ref{fig:convergence}) that beyond a distance of  $R_c = \SI{6}{\angstrom}$ the physical interactions are converged to less than $\SI{1}{\milli \electronvolt}$/atom, which is typically the target accuracy of the HDNNP fit with respect to the reference method.

\begin{table}[h!]
\caption{Parameters for the atom-centered symmetry functions~\cite{doi:10.1063/1.3553717} of the atomic NNs for copper and zinc as defined in Eqns.~(\ref{eq:sym1}) (type 2, radial) and (\ref{eq:sym2}) (type 3, angular). For the radial functions we have used $R_s=0$, the cutoff of all functions is $R_{\text{c}} = \SI{6}{\angstrom}$. For each specified parameter set all possible element combinations are considered. There are two radial and three angular functions per parameter set, resulting in 88 input neurons for each atomic NN.}
\begin{tabular}{cccrr}
  \centering
Number &Type  & $\eta$ ($\si{\bohr^{-2}}$)  & $\lambda$ & $\zeta$   \\ 
\hline
1   & 2    & 0.0010   & -          & -         \\
2   & 2    & 0.0200    & -          & -         \\
3   & 2    & 0.0350   & -          & -         \\
4   & 2    & 0.1000     & -          & -         \\
5   & 2    & 0.4000     & -          & -         \\
6   & 3    & 0.0001  & 1          & 1         \\
7   & 3    & 0.0001  & -1         & 2         \\
8   & 3    & 0.0030   & -1         & 1         \\
9   & 3    & 0.0030   & -1         & 2         \\
10  & 3    & 0.0080   & -1         & 1         \\
11  & 3    & 0.0080   & -1         & 2         \\
12  & 3    & 0.0080   & 1          & 2         \\
13  & 3    & 0.0150   & 1          & 1         \\
14  & 3    & 0.0150   & -1         & 2         \\
15  & 3    & 0.0150   & -1         & 4         \\
16  & 3    & 0.0150   & -1         & 16        \\
17  & 3    & 0.0250   & -1         & 1         \\
18  & 3    & 0.0250   & 1          & 1         \\
19  & 3    & 0.0250   & 1          & 2         \\
20  & 3    & 0.0250   & -1         & 4         \\
21  & 3    & 0.0250   & -1         & 16        \\
22  & 3    & 0.0250   & 1          & 16        \\
23  & 3    & 0.0450   & 1          & 1         \\
24  & 3    & 0.0450   & -1         & 2         \\
25  & 3    & 0.0450   & -1         & 4         \\
26  & 3    & 0.0450   & 1          & 4         \\
27  & 3    & 0.0450   & 1          & 16        \\
28  & 3    & 0.0800    & 1          & 1         \\
29  & 3    & 0.0800    & -1         & 2         \\
30  & 3    & 0.0800    & -1         & 4         \\
31  & 3    & 0.0800   & 1          & 4        
\end{tabular}
\label{tab:fit_paras}
\end{table}
For a binary system like brass, all element combinations need to be considered in the symmetry functions. Therefore, for each parameter set in Table~\ref{tab:fit_paras} there are two radial functions for each element of the central atom, one for copper neighbors and one for zinc neighbors. For the angular functions, any element combination is possible for the two neighboring atoms $j$ and $k$ resulting in three angular functions for each central element and parameter set.
It is important to note that all symmetry functions depend simultaneously on the positions of all atoms inside the cutoff sphere such that they are many-body functions. For a detailed description of symmetry functions and their properties we refer the interested reader to Ref. \citenum{doi:10.1063/1.3553717}.

\subsection{Generation of the Reference Structures} \label{sec:construct}

\begin{figure}[htb]
    \centering
\includegraphics[width=0.45\textwidth]{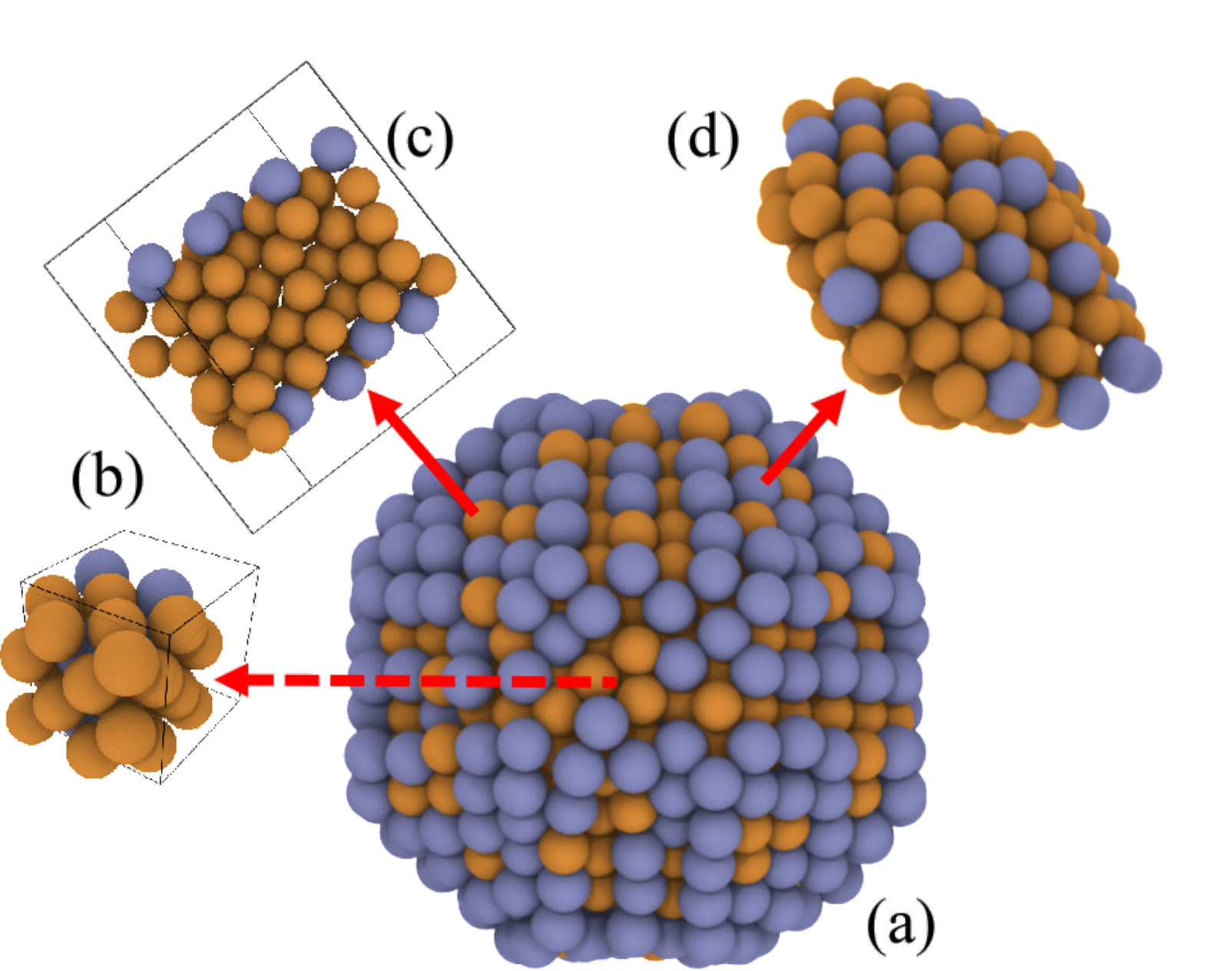}
 \caption{Schematic representation of the different types of atomic environments present in large brass nanoparticles (a). 
Because of the cutoff radii the environments can be represented by smaller bulk (b), surface (c) and cluster (d) calculations. All atomic visualizations in this work have been generated with the software OVITO~\cite{P5515}.
}
\label{fig:decompose}
\end{figure}

The HDNNP method relies on a decomposition of the system into local geometric motifs defined by the chosen cutoff radii of the atomic environments. Therefore, the same motifs that are found in very large systems can also be represented in much smaller bulk-, surface- and cluster structures as shown schematically in Fig.~\ref{fig:decompose}. In these systems, the values of the ACSFs describing the environment of a given central atom - or even of a  set of atoms - are thus identical to the corresponding values in the full system, which allows us to use small systems to train HDNNPs that are then transferable to large nanoparticles.

To ensure that the HDNNP is able to describe all important parts of the PES, the atomic environments in the training systems need to cover all structures that can emerge in the intended large-scale simulations. As the element distributions in nanoparticles can be very inhomogeneous, which we have also confirmed in our simulations~\cite{JanPaper2}, we have included structures with much higher global zinc atom fractions than 40$\%$, which forms the boundary of the $\alpha$-brass regime (see Fig.~\ref{fig:composition_env}).

We have started from an initial data set of 2,400 periodic $\alpha$-brass bulk structures that has been generated with the help of the Atomic Simulation Environment \cite{ase-paper} (ASE) library for Python. This set has then been extended step by step by additional bulk, surface and cluster structures.
For the terrace-like features of the nanoparticles resembling flat surfaces, we have generated slabs with Miller indices $(\text{hkl})$ up to $\text{h,k,l}=4$, and a thickness of at least the cutoff radius of the symmetry functions of the HDNNP. This ensures that atoms in the top and bottom layers of the slab do not lie within each others cutoff radius. However, arbitrary steps, edges, and kinks, or even more complex structural features, cannot be represented in slabs conveniently as the resulting supercells would become too large. More complex surface features of the nanoparticles are thus described by clusters, which are cut from the nanoparticles centered at the atoms of interest. Specifically, to extract new environments we start from large nanoparticles containing 1,103 atoms. From these nanoparticles all atoms within a radius of $\SI{12}{\angstrom}$ around an atom of interest are included in the extracted clusters. Note that in the present work we use the term cluster for rather small structures that are used to carry out DFT calculations, while the term nanoparticle is reserved for larger systems, which are only accessible by the HDNNP. The same procedure of extracting local clusters can also be applied to selected bulk-like environments, either from the interior of the nanoparticles or even from large bulk systems. 
The typical number of atoms in the obtained clusters with a maximum diameter of 24~\AA{} varies between 179 and 344, which is still affordable for reference DFT calculations. We did not include smaller clusters in the training set as these systems often represent open-shell systems with complex spin-multiplicities that can be difficult or even impossible to describe by a single potential energy surface. Further, small clusters would in many cases still be accessible by direct electronic structure calculations, while in the present work we focus on systems clearly beyond the realm of DFT, which are metallic and thus do not exhibit a complex electronic structure.

Because of the high computational costs of the reference DFT calculations and since this set of clusters might contain unnecessary redundant information, a selection procedure has to be implemented to identify those structures which add relevant new information to the data set, before the DFT calculations are carried out. To assess the relevance of a new candidate environment we rely on a procedure using only HDNNP predictions to identify missing important atomic environments~\cite{P3114}. For this purpose, we compare the predictions for the energies and forces for the clusters of interest using several different preliminary HDNNPs, which have been constructed using the same yet incomplete data set. Since HDNNPs have a very flexible functional form that is not based on physical considerations, different HDNNPs will provide similar energies and forces for a given atomic environment only if it is not too different from the environments which are already present in the data set and have been used in the training. On the other hand, if the predictions of several HDNNPs deviate substantially from each other, this is an indication that the candidate structure is too far from the known training data to be described reliably. In this case a DFT calculation should be carried out and the structure should be added to the reference set. This approach has the advantage that DFT calculations are only required for those structures that are really important, while they are not needed for the assessment if an atomic environment is sufficiently well described. The HDNNP parameters are then refined using this extended data set, and the procedure is repeated iteratively until the potential is reliable for all visited configurations. 

To ensure that this self-consistent generation of the reference data covers the structures that are needed for the intended applications, the candidate structures are identified under the same simulation conditions.
Here, to sample different configurations of bulk structures, slabs and clusters, we carry out molecular dynamics (MD) simulations with a timestep of $\SI{1}{\femto \second}$ in the $(NVT)$ and $(NpT)$ ensembles
using Nos\'{e}-Hoover chain thermostats\cite{Nose1984,P2756} in combination with Metropolis Monte Carlo exchange moves in the Semi Grand Canonical ensemble (SGCE). Details of these simulations can be found in Ref.~\citenum{JanPaper2}.
The simulations are performed with LAMMPS\cite{Plimpton1995} (version Aug. 2017) including an extension for HDNNPs \cite{Singraber2019}. In addition, a series of molten brass and copper structures was generated by performing MD simulations using preliminary versions of the HDNNP at a temperature of $\SI{1400}{\kelvin}$.
Including the energies and forces of these high temperature structures structures in the training set allows application of the HDNNP to molten brass structures with arbitrary atom coordinates.

\section{Results}\label{sec:results}

\subsection{Construction of the HDNNP}

\begin{figure}[h!]
         \centering
          \includegraphics[width=0.5\textwidth]{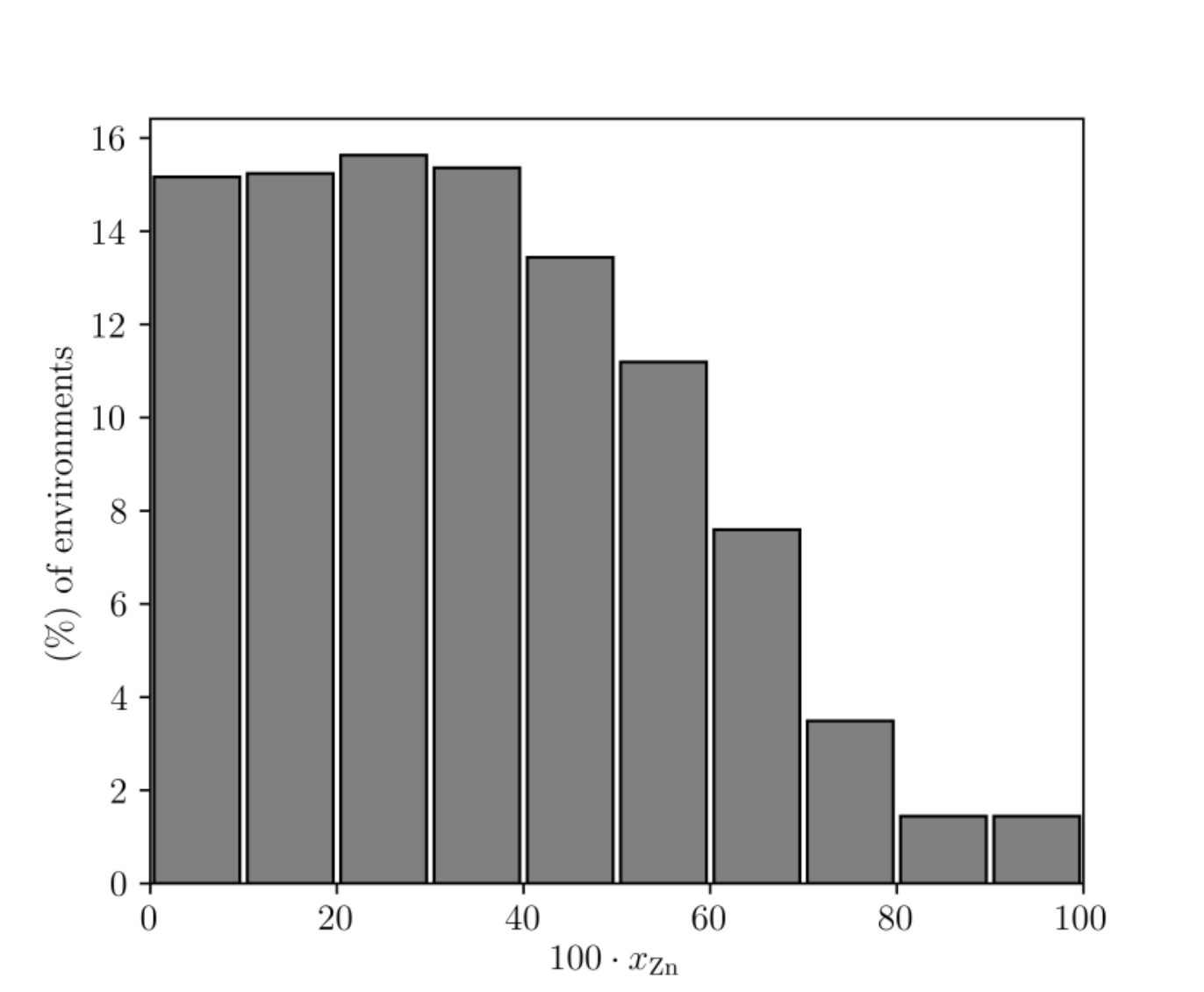}
         \caption{
         Chemical composition of the 2,500,000 atomic environments in the training set. Shown is the histogram of the environments as a function of the zinc atom fraction $x_{\text{Zn}}$.
         }
    \label{fig:composition_env} 
\end{figure}

To identify the optimum architecture for the atomic NNs, which for simplicity are kept identical for both elements, we tested different numbers of hidden layers and nodes per layer. 
The weight parameters, which have been initialized according to the Nguyen and Widrow scheme \cite{Nguyen}, are iteratively optimized to reproduce the DFT energies and forces using the HDNNP program RuNNer \cite{Jorg,P5128,P4444} 
employing the global extended Kalman filter algorithm \cite{Kalman1960, blank1994}.  
In total 53,841 reference structures have been generated including 4,009 brass clusters, 8,492 molten brass bulk structures, 8,964 copper slabs, and 16,878 brass slabs. Additionally, 5,377 copper bulk structures and 10,121 brass bulk structures have been included. This data set contains 2,967,780 atomic environments, 53,841 total energies and 8,903,340 force components. The data has been split into a training set to optimize the NN weight parameters (85~\%) and an independent test set (15~\%) to estimate the accuracy for structures not included in the training. The data covers essentially all possible zinc atom fractions in the atomic environments, as is shown in Fig.~\ref{fig:composition_env} for the 2,521,617 atomic environments of the training set. 2,040,143 of these environments refer to copper atoms, 481,474 to zinc atoms, and we note that the compositions of the environments of both elements are essentially the same. The ranges of values for the energies and force components to be fitted have a width of about 2~eV/atom and $\SI{15}{\electronvolt \per \angstrom}$, respectively. 

\begin{figure*}[htb]
    \centering
\includegraphics[width=0.9\textwidth]{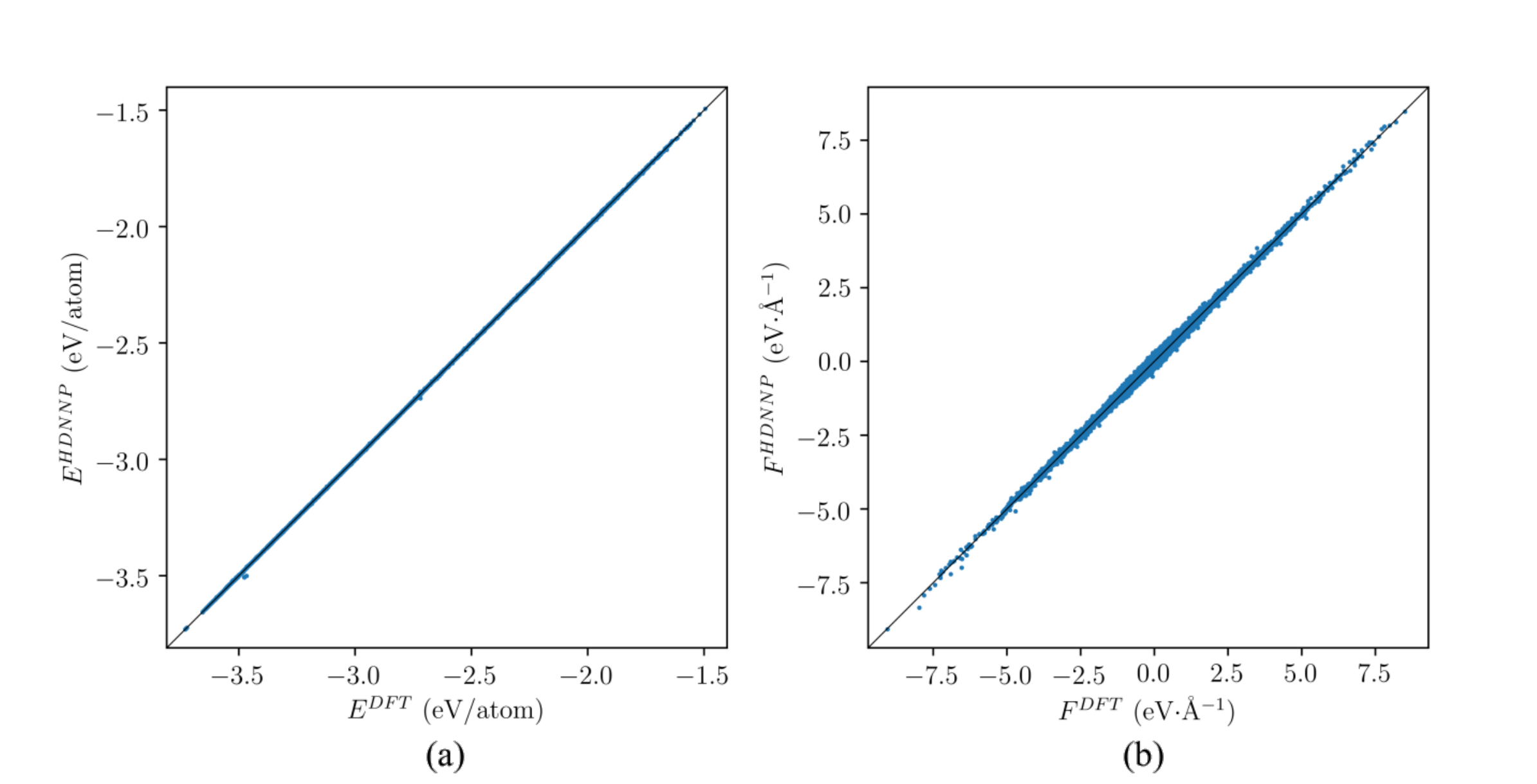}
\caption{Correlation plots for the HDNNP and DFT energies (a) and atomic force components (b) in the test set. The HDNNP representation is in very good agreement with DFT over the whole range of values.}
\label{fig:accuracy_summary}
\end{figure*}

Table~\ref{tab:arch} shows the accuracy of the HDNNP for different neural network architectures. Based on the root mean squared errors (RMSE) of the energies $E^{\text{RMSE}}$ and forces $F^{\text{RMSE}}$ of the training and the test set we found that two hidden layers containing 20 nodes each offers the best compromise between accuracy and size of the architecture.
The final RMSEs for this architecture are $\SI{1.8}{\milli \electronvolt/}$atom for the energies and $\SI{39.5}{ \milli \electronvolt \per \angstrom}$ for the forces in the training set. The RMSEs of the energies and forces of the test structures are $\SI{1.7}{\milli \electronvolt /}$atom and $\SI{38.7}{\milli \electronvolt\per \angstrom}$ respectively, which is very similar to the accuracy for the training set, indicating the absence of overfitting. Fig.~\ref{fig:accuracy_summary} shows the correlation plot of the HDNNP and DFT energies and forces for the test set.

\begin{table}[htb]
\center
\caption{Comparison of the energy and force RMSE values of the training and test sets for different NN architectures after 20 epochs of training. Atomic NNs with two hidden layers and 20 nodes in each layer (20-20) have been selected for the final HDNNP for further analysis (indicated in bold). 
}
\begin{tabular}{lcccc}
\hline
\multicolumn{1}{l}{} & \multicolumn{2}{c}{$E^{\text{RMSE}}$ ($\si{\milli \electronvolt}/\text{atom}$)} & \multicolumn{2}{c}{$F^{\text{RMSE}}$ ($\si{\milli \electronvolt \per \angstrom}$)} \\
NN & Training & Test & Training & Test \\ \hline
10-10 & 1.88 & 1.76 & 41.7 & 41.1 \\
15-15 & 1.82 & 1.71 & 40.9 & 40.3 \\
15-15-15 & 1.84 & 1.75 & 41.3 & 40.7 \\
\textbf{20-20} & \textbf{1.77} & \textbf{1.67} & \textbf{39.5} & \textbf{38.8} \\
20-20-20 & 1.75 & 1.66 & 39.6 & 39.0
\end{tabular}
\label{tab:arch}
\end{table}

\subsection{Validation of the HDNNP for the Bulk} 
\label{sec:first_test}
\label{sec:validation}

\begin{table}[htb]
\caption{Equilibrium lattice constant $a$ of bulk copper  
as well as cohesive energies of bulk copper and three ordered $\alpha$-brass phases shown in Fig.~\ref{fig:configs}.}
\begin{tabular}{lccc}
Bulk Property                              & HDNNP                   & DFT & Experiment  \\
\hline
$a$ fcc Cu ($\si{\angstrom}$)                    & 3.63                  & 3.63          & 3.61  Ref.~\citenum{Davey1925}                             \\
$E_{\text{coh}}$ fcc Cu ($\si{\electronvolt}$/atom)       & -3.488                & -3.487        & -3.49 Ref.~\citenum{Kittel2005}   \\ 
$E_{\text{coh}}$ $L_{12}$ ($\si{\electronvolt}$/atom)                         & -2.957 & -2.954    & -   \\
$E_{\text{coh}}$ $DO_{23}$ ($\si{\electronvolt}$/atom)                        & -2.956 & -2.955    & -   \\
$E_{\text{coh}}$ $LPS_3$ ($\si{\electronvolt}$/atom)                           & -2.955 & -2.953    & -  \\
\end{tabular}
\label{tab:bulk1}
\end{table}

For validating the constructed potential we start by comparing the HDNNP and DFT results for different structural and energetic properties of bulk copper and $\alpha$-brass. First we determined the equilibrium lattice constant and the cohesive energy of pure fcc copper using the HDNNP. The obtained results (see Tab.~\ref{tab:bulk1}) are in excellent agreement with DFT and experiment. The small deviation in the lattice constant of about 0.02~\AA{} for the HDNNP with respect to experiment can be attributed to the employed exchange-correlation functional since a HDNNP representation of the PES can only be as accurate as the reference method.

In the remaining part of this section we will assess the accuracy of the HDNNP by systematically testing its performance for brass structures.
First, we replace a single copper atom in a bulk cell by zinc, then we investigate structures formed for a zinc atom fraction of 25~\% and finally we address arbitrary compositions.

As a first step, we replace a single atom in a $2 \times 2 \times 2$ bulk copper cell containing 32 atoms by a zinc atom, which will be important for simulations e.g. in the semi-grand canonical ensemble~\cite{JanPaper2}, and we fully relax the system before and after the atom exchange. 
Using DFT we find an energy increase of $\Delta E_{\text{Cu} \rightarrow \text{Zn}}^{\text{DFT}} =\SI{67.4}{\milli \electronvolt}$, while the corresponding value of the HDNNP is $\Delta E_{\text{Cu} \rightarrow \text{Zn}}^{\text{HDNNP}}= \SI{66.2}{\milli \electronvolt}$. The difference of only about $\SI{1.3}{\milli \electronvolt}$ is  below the total energy RMSE of the HDNNP, but this comparison has to be made with care, because the energy RMSE is given per atom resulting potentially in a much larger uncertainty of the energy for a many-atom system like a bulk supercell, while the energy increase upon the investigated atom exchange is the global energy change of the full structure.
However, energy differences like in the present test usually exhibit much smaller errors than total energies due to error cancellation between those parts of the system that are not affected by the substitution, which is clearly confirmed in the present case. As a result we find that the agreement between the HDNNP and DFT for small compositional changes is very good.

\begin{figure*}[htb]
\centering
\includegraphics[width=0.7\textwidth]{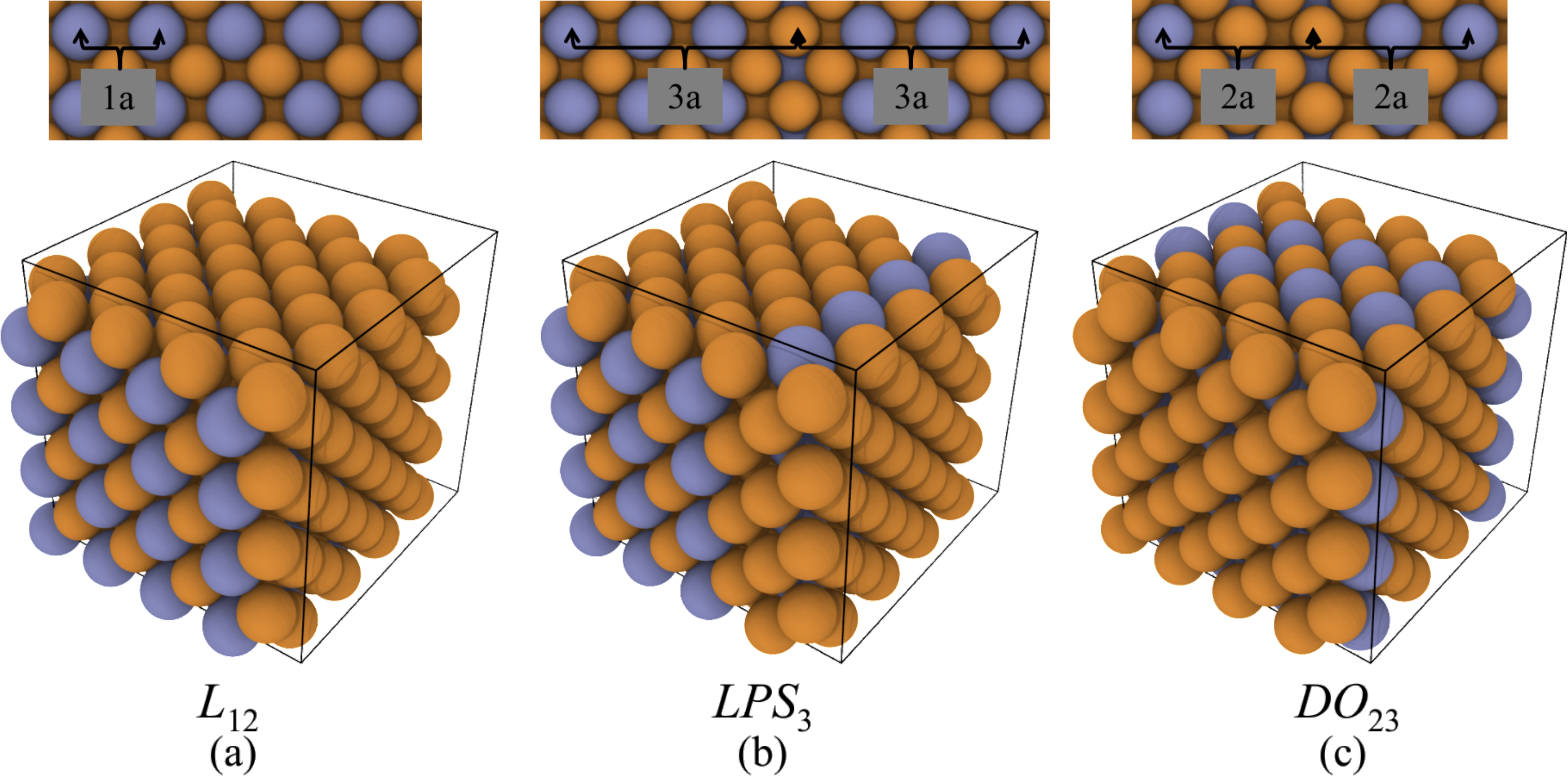}
\caption{Lowest energy configurations 
corresponding to the $L_{12}$, $LPS_3$, $DO_{23}$ bulk phases of  $\alpha$-brass for a composition of $\text{Cu}_{0.75}\text{Zn}_{0.25}$ found in Metropolis Monte Carlo (MMC) simulations of a $N=256$ atom supercell employing the HDNNP. These structures are in excellent agreement with Ref.~\citenum{p2119} . 
}
    \label{fig:configs}
\end{figure*}

To further investigate the exchange of copper and zinc atoms, we use the HDNNP in Metropolis Monte Carlo simulations of a bulk brass supercell containing 256 atoms with a fixed zinc atom fraction $x_{\text{Zn}} = 25 \%$, allowing for copper-zinc position switches within the system as the only trial moves. In these simulations, we find three structures, $L_{12}$, $DO_{23}$, and $LPS_3$, differing only in the relative positions of the zinc layers (see Fig.~\ref{fig:configs}), which are the lowest energy configurations of $\alpha$-brass at this concentration as reported previously~\cite{p2119}. The HDNNP and DFT cohesive energies of these structures (see Tab.~\ref{tab:bulk1}) are extremely similar spanning a range of only 2 meV/atom and thus cannot be distinguished with confidence even at the DFT level. Still, the overall agreement between DFT and the HDNNP is remarkable taking into account that the total energy change when replacing even a single copper atom by zinc is about two orders of magnitude larger.

Finally, to validate the HDNNP for arbitrary compositions we created a set of $2 \times 2 \times 2$ supercells of bulk brass containing 32 atoms.
In total thirteen different zinc concentrations including the case of pure copper are generated covering the complete $\alpha$-brass regime.
Note that the bulk cells with the two lowest zinc concentrations only contain zero and one zinc atom, respectively, and thus there is just one possible configuration due to symmetry. For all other concentrations we generate one hundred configurations, differing in the occupation of different sites by copper or zinc. 
Each of the in total $1,102$ structures has been fully relaxed.

\begin{figure}[h!]
         \centering
          \includegraphics[width=0.45\textwidth]{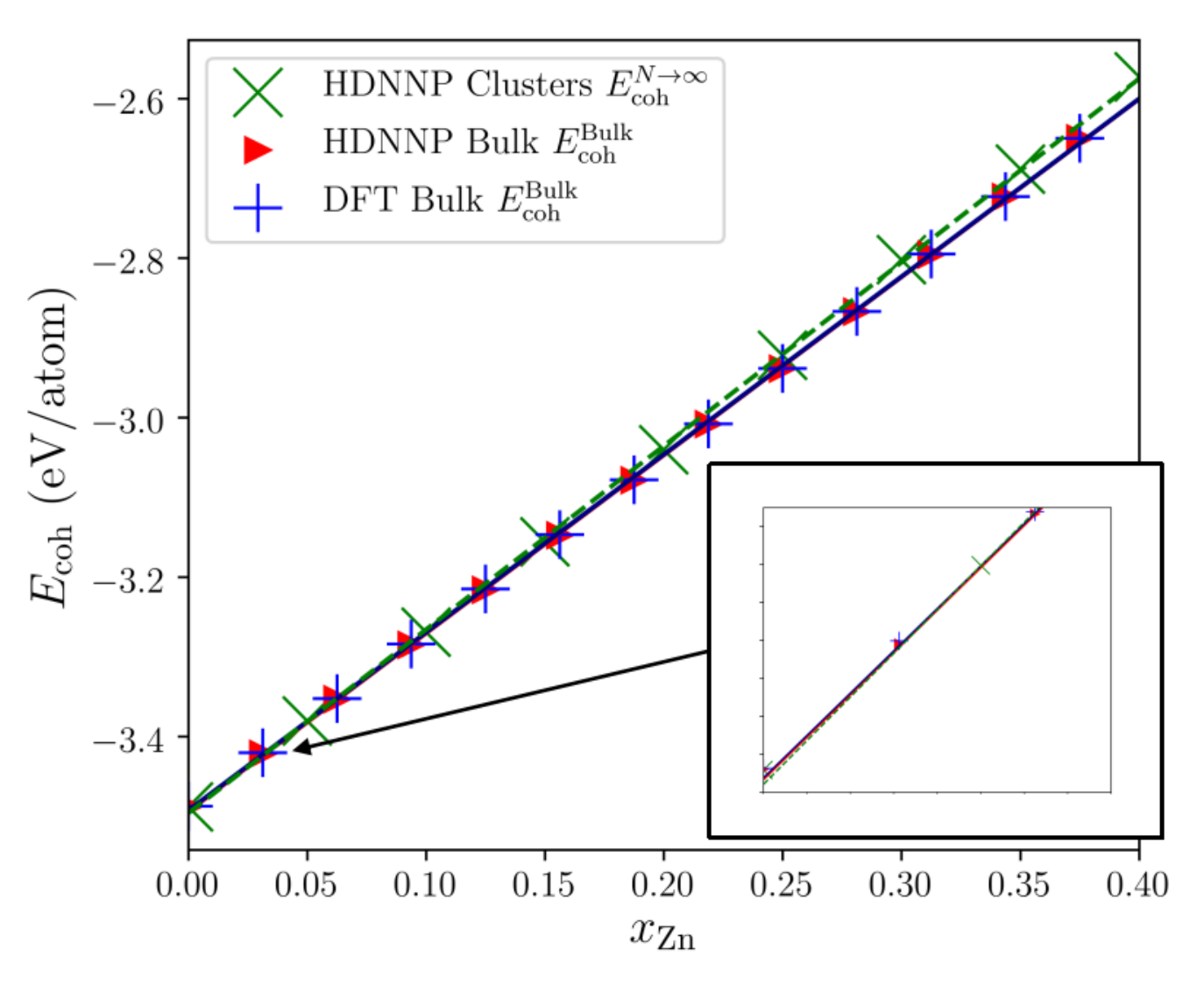}
 \caption{
 Cohesive energies $E_{\rm coh}$ for 32-atom relaxed bulk cells of $\alpha$-brass with varying zinc atom fraction $x_{\rm Zn}$ obtained from the HDNNP (red) and DFT (blue) and linear regressions to the data. $E_{\rm coh}^{N \rightarrow \infty}$ (green dashed) refers to extrapolated data for clusters using intercept values of linear regressions as
 explained in Fig.~\ref{fig:size_cohe}b in sec.~\ref{sec:cohe_brass}. 
 } 
    \label{fig:ecoh_inf}
\end{figure}

\begin{figure}
    \centering
\includegraphics[width=0.45\textwidth]{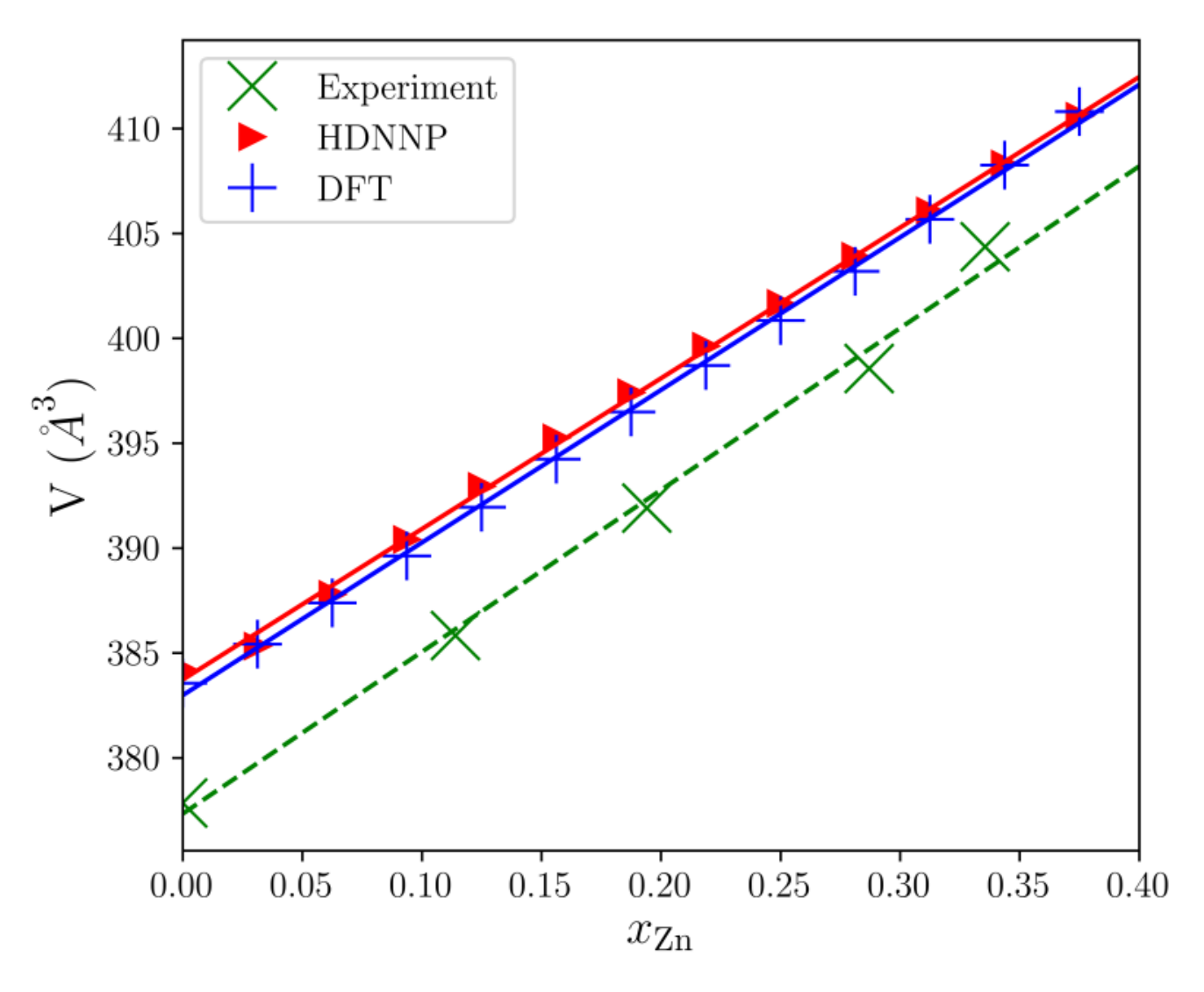}
    \caption{
    Volumes of 32 atom $\alpha$-brass bulk supercells as a function of the zinc atom fraction $x_{\text{Zn}}$ computed by the HDNNP (red) and DFT (blue). For all but the two lowest zinc atom fractions 100 different configurations have been averaged for both methods. For comparison, also experimental data\cite{Davey1925} is given (green dashed line). 
    }
    \label{fig:expansion}
\end{figure}

We observe that the cohesive energy increases to less negative values when the zinc concentration is increased, and we generally find that $E_{\text{coh}}^{\text{Cu}}  < E_{\text{coh}}^{\text{CuZn}} < E_{\text{coh}}^{\text{Zn}}$.
Essentially, the HDNNP and DFT predict a linear increase of the cohesive energy $E_{\text{coh}}$ of the 32 atom cell of $\alpha$-brass as a function of the zinc atom fraction $x_{\text{Zn}}$. The same is also true for the volume $V$ of the cell.
The average cell expansion coefficient and slope of the composition dependent cohesive energy are calculated using linear regression with respect to average values for each of the thirteen sampled zinc atom fractions.
The slope of the cohesive energy $E_{\text{coh}}$ as a function of $x_{\text{Zn}}$ is $\SI{69.75}{\milli \electronvolt}$ per atom in the system
(HDNNP) and $\SI{69.69}{\milli \electronvolt}$ per atom (DFT). These results are in good agreement with the value for the increase in cohesive energy for exchanging one copper atom with a zinc atom in a 32 atom copper supercell as discussed earlier in this section.
This underlines the very good agreement between the HDNNP and DFT energetics for compositional changes over the complete range of zinc contents reaching from $x_\text{Zn}=0.0$ to $x_\text{Zn}=0.4$ (see Fig.~\ref{fig:ecoh_inf}). The HDNNP result for the volume expansion coefficient of $\SI{2.25}{\angstrom^3}$ per atom, i.e. the slope of the volume as a function of the zinc atom fraction shown in Fig.~\ref{fig:expansion},
is also in excellent agreement with the DFT value of $\SI{2.27}{\angstrom^3}$ per atom corresponding to an error of only about $1 ~\%$.
The corresponding experimental slope of $\SI{2.41}{\angstrom^3}$ per atom, which has been estimated from the lattice constants given in Ref.~\citenum{Davey1925}, deviates by approximately $7 ~\%$, which we consequently attribute to the underlying exchange correlation functional, since the HDNNP is constructed to reproduce the reference method.

\subsection{Surface Energies and Wulff Construction}\label{sec:wulff}

A method to determine the equilibrium shape of nanoparticles is the Wulff construction \cite{Wulff1901}, which relies on an accurate description of the surface energies. The original Wulff construction is only applicable to large nanoparticles of a single element like copper, and the extension to alloys is not straightforward, because the surface energy does not only depend on the geometry of the surface but also on its composition and the specific occupations of the lattice sites. Only recently, the Wulff construction has been generalized to alloy nanoparticles~\cite{Ringe2011a}.

To obtain initial structural models for brass nanoparticles, we use the Wulff construction and the surface energies $\sigma_{\text{(hkl)}}$ of copper. We then cut nanoparticles with this shape and the desired size from the fcc copper bulk lattice. Finally, a given fraction of randomly selected copper atoms is replaced by zinc atoms according to the zinc atom fraction of interest. Any deviation of the particle shape from the copper-based Wulff construction is then obtained by equilibrating the system in subsequent structural relaxations and Monte Carlo simulations with swaps as trial moves.

\begin{figure}[htb]
\centering
         \includegraphics[width=0.4\textwidth]{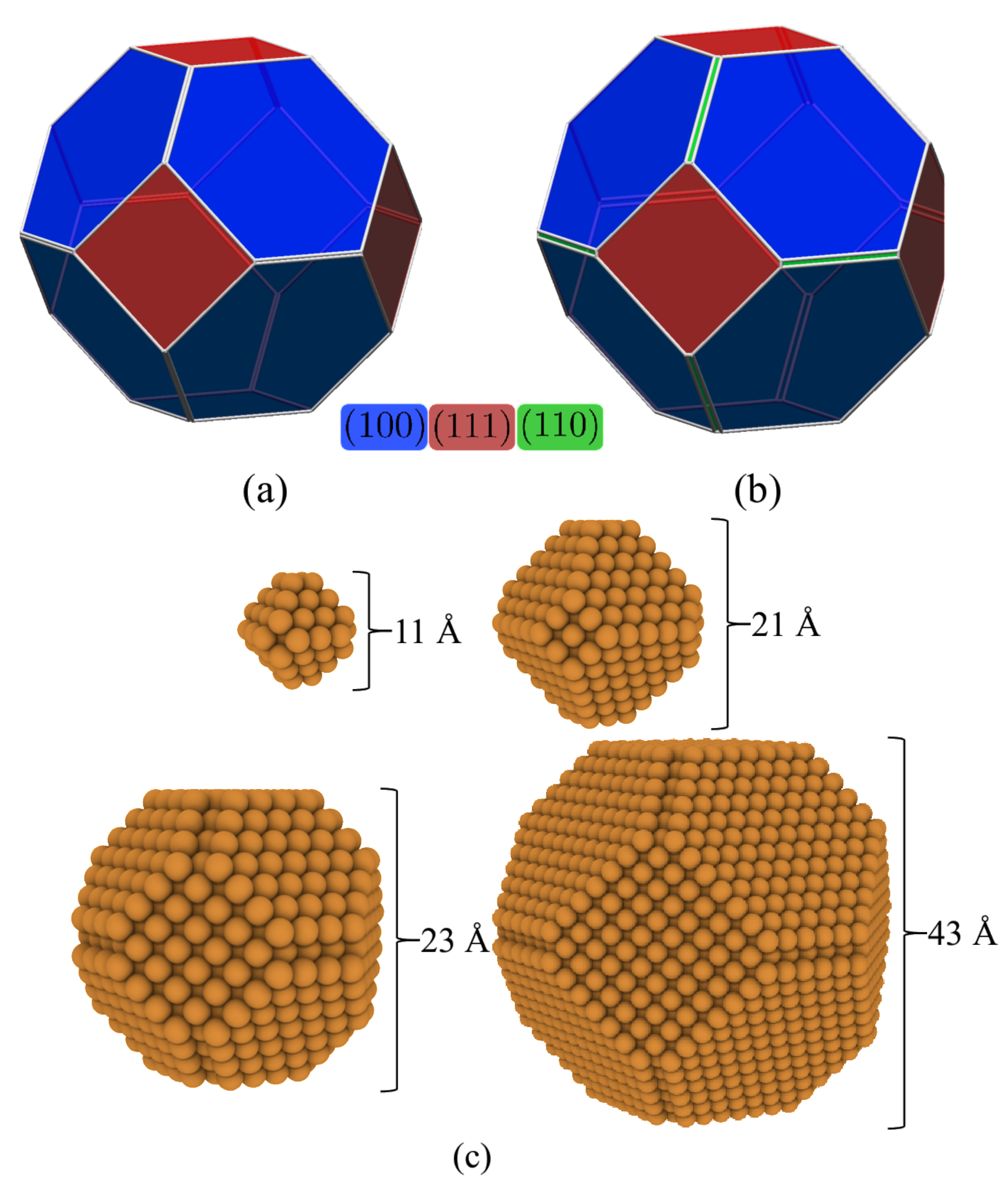}
\caption{Wulff shapes \cite{weber} of copper nanoparticles obtained from the HDNNP (a) and DFT (b) constructed\cite{ase-paper} using the surface energies in Tab.~\ref{tab:values_surface}. Panel (c) shows the derived copper structures containing $N=79,~459,~1,103, \text{and} ~4,897$ atoms, respectively.}
    \label{fig:wulff_const}
\end{figure}

The surface energies of the most stable (111), (110), and (100) surfaces of copper have been calculated using the method by Fiorentini \cite{Fiorentini1996}, and the results are given in Table \ref{tab:values_surface}.
The HDNNP surface energies are in the same energetic order
$\sigma_{(1 1 1)} < \sigma_{(100)} < \sigma_{(110)}$ as the corresponding DFT values with only very small deviations of at most $\SI{1}{\milli \electronvolt \per \angstrom \squared}$, and the deviations with respect to experiment can be ascribed to both, experimental uncertainties as well as limitations of current GGA functionals. 
Consequently, the HDNNP and DFT Wulff shapes of copper nanoparticles shown in Fig.~\ref{fig:wulff_const} are extremely similar.  Because of the finite number of atoms and the resulting discretization, slight differences between clusters based on the DFT or HDNNP surface energies only emerge for systems containing more than approximately 3,800 atoms, and in that case we have based our initial structural models on the HDNNP surface energies.
Specifically, here we will investigate brass nanoparticles containing 79 (diameter $\approx \SI{1}{\nano \meter}$), 459 ($\approx \SI{2}{\nano \meter}$), 1,103 ($\approx \SI{2.5}{\nano \meter}$) and 4,897 atoms ($\approx \SI{4}{\nano \meter}$) (see Fig.~\ref{fig:wulff_const}) to carry out simulations and to investigate the size dependence of the results.

\begin{table*}
\centering
\caption{Surface energies $\sigma_{\text{(hkl)}}$ of the most stable copper surfaces obtained using the HDNNP and DFT. Experimental numbers are also given for comparison. An experimental value for the (100) surface energy was estimated based on the surface energy of the (111) surface\cite{Vitos1998} and the ratio $\sigma_{\text{(111)}}/\sigma_{\text{(100)}} = 0.994$ \cite{doi:10.1063/1.114455}.
}
\begin{tabular}{lccc}
Surface & $\sigma^{\text{HDNNP}}_{\text{(hkl)}}$($\si{\milli \electronvolt \per \angstrom \squared}$)  & $\sigma^{\text{DFT}}_{\text{(hkl)}}$($\si{\milli \electronvolt \per \angstrom \squared}$) & $\sigma^{\text{Exp.}}_{\text{(hkl)}}$($\si{\milli \electronvolt \per \angstrom \squared}$)   \\
\hline
(100)                 & 92        & 92        & $\approx 113$ \cite{doi:10.1063/1.114455, Vitos1998}     \\
(110)                 & 99        & 98        & 114\cite{Fishman}                \\
(111)                 & 82        & 81        & 112\cite{Vitos1998}                 \\
\end{tabular}
\label{tab:values_surface}
\end{table*}

\subsection{Validation of the HDNNP for Clusters}

In this section we investigate the accuracy of the HDNNP for clusters starting with pure copper followed by clusters containing increasing zinc contents. Note that the applicability of the HDNNP is not restricted to atomic positions corresponding to an fcc lattice. Therefore, the validation and comparison with DFT for structures deviating substantially from the fcc lattice, e.g. at high temperatures, is of particular interest.

\subsubsection{Melting of Copper Clusters}
\label{sec:size_scaling}

In general the melting temperature $T_{m}$ of a nanoparticle depends on its shape\cite{Cui2017a} and it is well-known that $T_{m}$  of spherical nanoparticles increases with the radius $R$ and the number of atoms $N$ following the relation \cite{Nanda2009,Kart2014,Johnston, Wilcoxon} 

\begin{align}
         T_{m}(R) = T_{\text{m}}^{\text{Bulk}} \left( 1 - \frac{k}{R} \right) ~. 
         \label{eq:melt}
\end{align}

$T_{\rm m}^{\rm Bulk}$ is the corresponding melting temperature of the bulk material and $k$ is a constant.
Here, we will apply this relation not only to spherical but also to Wulff-shaped copper nanoparticles assuming that the Wulff nanoparticles can be effectively treated as a sphere with volume $V= \frac{4}{3} \pi \cdot R_{\text{eff}}^3$.
The effective radius $R_{\text{eff}}(N)$ for a given number of atoms $N$ allows to express the relation for the melting point (see Eq.~(\ref{eq:melt})) in terms of $N$ using $\rho=N/V$ as follows,
\begin{align}
    T_{m} (N) &=&     
    T_{\text{m}}^{\text{Bulk}}\left[1  -   
    \underbrace{k
    \left( \frac{3N}{4 \rho \pi}   \right)^{-\frac{1}{3}}
    }_{1/R_{\text{eff}}}\right] = T_{\text{m}}^{\text{Bulk}} - \frac{k^{\prime}}{N^{\frac{1}{3}}}  \label{eq:melt_law} ~,
\end{align}
\\
i.e., there should be a $T_{m} \propto N^{-\frac{1}{3}}$ dependence of the melting temperature on the number of atoms of the nanoparticle.

Many important thermodynamic quantities like the cohesive energy $E_{\text{coh}}$, the melting enthalpy $H_{m}$, and the heat of segregation $Q_{m}$ follow the same scaling law\cite{Cui2017a}. Therefore, $T_{m} \propto E_{\text{coh}} \propto H_{m} \propto Q_{m} \propto N^{-\frac{1}{3}}$, and it is an important test to investigate if the HDNNP can reproduce this scaling law correctly.
Note that according to Eq.~(\ref{eq:melt_law}) very large nanoparticles with $N \rightarrow \infty$ approach the melting temperature of bulk copper $T_{\text{m}}^{\text{Bulk}}$.

\begin{figure*}[h!]
         \centering
          \includegraphics[width=0.9\textwidth]{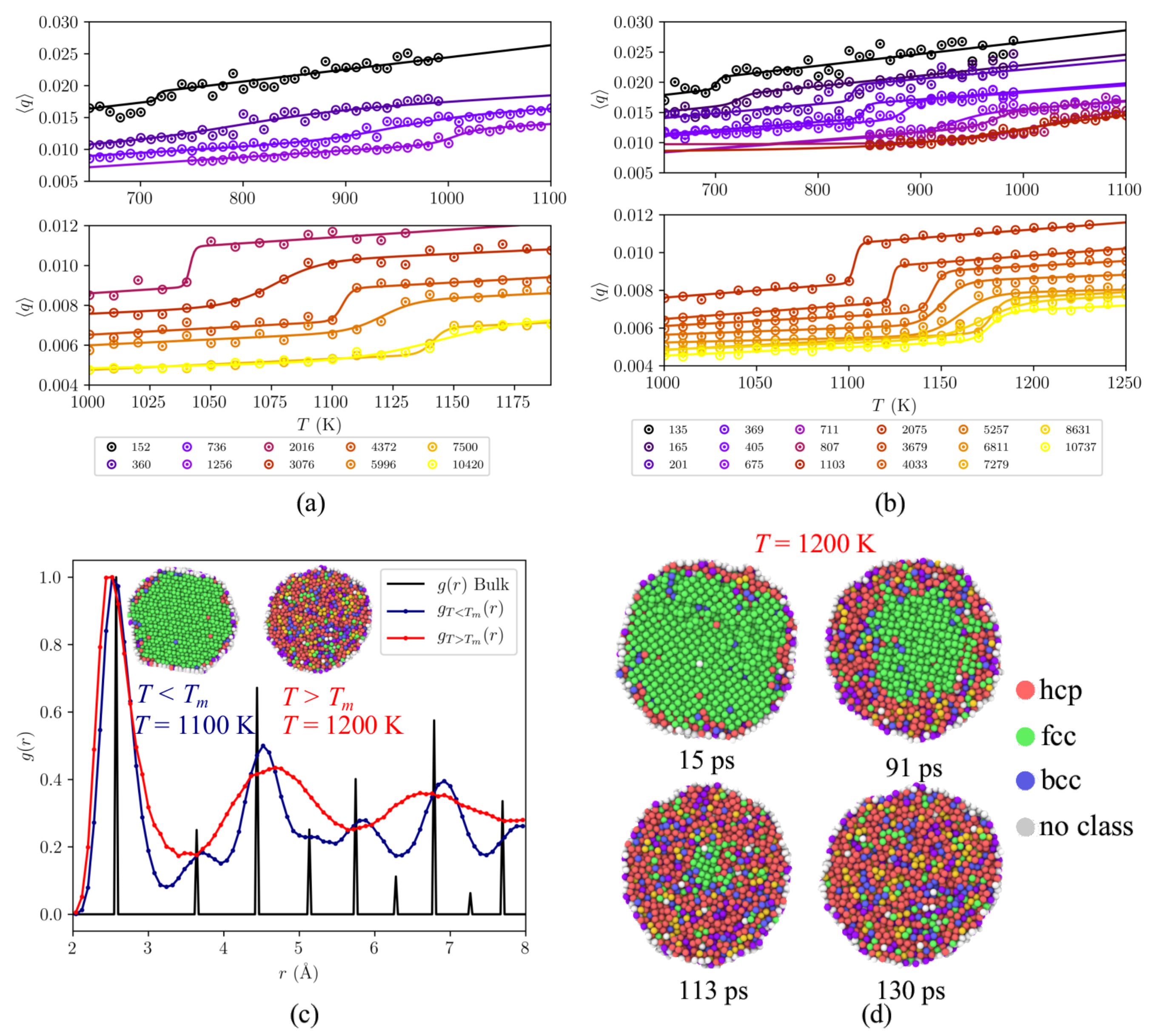}
         \caption{
         Global Lindemann index $\langle q \rangle$ defined in Eq.~(\ref{eq:lindemannlocal}) as a function of temperature $T$ for copper nanoparticles containing different numbers of atoms $N$ with initial spherical shape (a) or Wulff-shape (b), respectively.
         The lines in (a) and (b) are fits to the data points using the empirical function for $\langle q \rangle$ defined in Eq.~(\ref{eq:empirical}).
         In (c) we show normalized radial distribution functions $g(r)$ for the $N=$ 10,737 atom copper nanoparticle at $T=\SI{1100}{\kelvin}$ and $T=\SI{1200}{\kelvin}$  as well as $g(r)$ for bulk copper. The melting temperature of this nanoparticle is $T_{m} = \SI{1176}{\kelvin}$. In (d) we show that the melting process starts at the surface and proceeds to the center of the nanoparticle. Two polyhedral template matching\cite{P5714} (PHTM) color maps show the local crystal order at these temperatures (green is fcc, red hcp, blue bcc, white no type). 
         } 
    \label{fig:lindemannPlots} 
\end{figure*}

First, we address pure copper clusters of varying size and estimate the melting temperatures using MD simulations in the $(NVT)$ ensemble based on the HDNNP. The temperature is controlled by Nos\'{e}-Hoover chain thermostats~\cite{Nose1984, Hoover1996} and we use an integration time step of $\Delta t = \SI{1.4}{\femto \second}$. At each temperature we compute trajectories with a total simulation time of $\SI{0.21}{\nano \second}$. Clusters of different sizes containing between 135 and 10,737 atoms have been investigated starting from two different initial shapes, one derived from a Wulff construction and the other one being approximately spherical. We observe the well-known behavior \cite{Kart2014} that melting of nanoparticles starts at the surface and proceeds to interior regions as showcased in Fig.~\ref{fig:lindemannPlots}d.
This process takes place on a much shorter time scale compared to the bulk crystal and typically only a few hundreds of $\si{\pico \second}$ of simulation time are necessary for melting, because surface atoms are less spatially constrained and thus have a higher mobility than atoms in the center of the nanoparticle or in the bulk.

For a better qualitative understanding of the melting process we compare the radial distribution functions $g(r)$ of a copper nanoparticle with $N=$ 10,737 atoms ($T_m$=1176 K) before and after melting
at $T=\SI{1100}{\kelvin}$ and $T=\SI{1200}{\kelvin}$ in Fig.~\ref{fig:lindemannPlots}c. We find that after melting the  peak at the second nearest neighbour distance of $\SI{3.63}{\angstrom}$ has vanished. This is consistent with the local crystal ordering assigned to every atom by polyhedral template matching \cite{P5714} (PHTM). We note that the non-zero $g(r)$ values in between the bulk peak positions are a consequence of the increased mobility of some under-coordinated surface atoms, which do not adopt fcc bulk like distances even below the melting temperature. 

For the analysis of the melting process we assign the local Lindemann index\cite{Lindemann1910} $q_{i}$ to each atom $i$, which is defined as

\begin{align}
    { q_{i} = 
    \frac{1}{N-1} \sum _{j\neq i}{\frac {\sqrt {\langle r_{ij}^{2}\rangle_{\text{MD}} -\langle r_{ij}\rangle_{\text{MD}} ^{2}}}{\langle r_{ij}\rangle_{\text{MD}} }}} ~,
    \label{eq:lindemannlocal} 
\end{align}

where we denote the average over all sampled configurations of an MD simulation with $\langle . \rangle_{\text{MD}}$.
The atomic values $q_i$ are essentially normalized standard deviations of all the inter-atomic distances $r_{ij}$ of a reference atom $i$ with all other atoms $j$ in a given geometry.
The sum in Eq.~(\ref{eq:lindemannlocal}) is computed for all atoms in the system corrected for double counting of the distances $r_{ij}$. 
The Lindemann index takes a value of 0 for systems with no mobility where the interatomic distances remain constant across the simulation, and increases as the atoms in the system become more mobile. Therefore, phase transitions like melting give rise to a notable change in the Lindemann index.

The global Lindemann index\cite{Lindemann1910} $\langle q \rangle$ is defined as the system average over all atomic values $q_{i}$.
It is often observed that the global Lindemann index of crystalline systems increases linearly with temperature until the melting point $T_{m}$ is reached, which is characterized by a sudden non-linear increase\cite{Zhang_2007}. 

We compute the global Lindemann index for a series of temperatures using temperature intervals of $\Delta T = \SI{10}{\kelvin}$. The results are shown in Figs.~\ref{fig:lindemannPlots}a,b. 
As a substantial change of the Lindemann index is indication of an order-disorder transition from a crystal to the liquid phase, we determine the melting temperatures by fitting
modified logistic sigmoid functions of the form
\begin{align}
\langle q(T) \rangle \approx \frac{x_{1}}{1+e^{-x_{2} (T-T_{m})}} +x_{3} \cdot T+x_{4}~,
\label{eq:empirical}
\end{align}
with parameters $T_{m},x_{1}, x_{2}, x_{3}$ and $x_{4}$. We then define the position of the turning point as given by the parameter $T_m$ in the sigmoid as the melting point of the nanoparticle. 
The resulting melting points $T_{m}$ are listed in Tab.~\ref{tab:empirical_paras}. 
The linear term $x_3 \cdot T+x_4$ in Eq.~(\ref{eq:empirical}) has been motivated by the observation that $\langle q \rangle$ can be approximated with good accuracy by a linear function for temperatures that are lower or higher than the melting point. We observe that the turning points of the Lindemann curves are shifted to higher temperatures for larger nanoparticles yielding the expected increase in melting temperature with nanoparticle size.

\setlength{\tabcolsep}{0.7pt}
\begin{table}[h!]
\begin{tabular}{c|c|c|c|c|c|c}
\multicolumn{6}{c}{Spherical Shape} \\
\hline
\multicolumn{1}{c|}{$N$} & \multicolumn{1}{c|}{$x_{1}$} & \multicolumn{1}{c|}{$x_{2}$} &\multicolumn{1}{c|}{$x_{3}$} & \multicolumn{1}{c|}{$x_{4}$} & \multicolumn{1}{c}{$T_{m}$} \\
\multicolumn{1}{c|}{} & \multicolumn{1}{c|}{$\times 10^{-3}$} & \multicolumn{1}{c|}{($\si{\per \kelvin} \times 10^{-1}$)} & \multicolumn{1}{c|}{($\si{\per \kelvin} \times 10^{-6}$)} & \multicolumn{1}{c|}{$\times 10^{-3}$} & \multicolumn{1}{c}{($\si{\kelvin}$)} \\
\hline
152 & 1.4 & 5.8 & 19.0 & 4.0 & 717 \\
360 & 3.6 & 0.2 & 9.6 & 4.3 & 792 \\
736 & 3.1 & 0.3 & 9.6 & 2.7 & 946 \\
1,256 & 2.5 & -0.9 & 10.5 & 2.4 & 1005 \\
2,016 & 2.0 & 10.4 & 87.9 & 0.6 & 1041 \\
3,076 & -2.2 & -1.1 & 5.5 & 4.2 & 1077 \\
4,372 & -1.6 & -5.2 & 6.6 & 1.5 & 1104 \\
5,996 & -1.5 & -1.4 & 5.9 & 1.6 & 1121 \\
7,500 & 1.3 & 3.6 & 5.6 & -0.8 & 1141 \\
10,420 & 1.9 & 0.6 & 2.9 & 1.9 & 1144 \\
\end{tabular}
\begin{tabular}{c|c|c|c|c|c|c}
\multicolumn{6}{c}{Wulff Shape} \\
\hline
\multicolumn{1}{c|}{$N$} & \multicolumn{1}{c|}{$x_{1}$} & \multicolumn{1}{c|}{$x_{2}$} &\multicolumn{1}{c|}{$x_{3}$} & \multicolumn{1}{c|}{$x_{4}$} & \multicolumn{1}{c}{$T_{m}$} \\
\multicolumn{1}{c|}{} & \multicolumn{1}{c|}{$\times 10^{-3}$} & \multicolumn{1}{c|}{($\si{\per \kelvin} \times 10^{-1}$)} & \multicolumn{1}{c|}{($\si{\per \kelvin} \times 10^{-6}$)} & \multicolumn{1}{c|}{$\times 10^{-3}$} & \multicolumn{1}{c}{($\si{\kelvin}$)} \\
135 & 1.9 & 4.5 & 19.6 & 5.2 & 701 \\
165 & -1.6 & -1.7 & 17.4 & 5.4 & 723 \\
201 & 2.7 & 4.1 & 15.4 & 4.1 & 834 \\
369 & 2.4 & -3.2 & 13.3 & 5.2 & 844 \\
405 & 3.0 & 1.4 & 11.9 & 3.4 & 886 \\
675 & -2.9 & -0.6 & 12.4 & 3.2 & 932 \\
711 & -2.5 & -0.9 & 13.3 & 2.3 & 973 \\
807 & -5.9 & -0.3 & 0.32 & 15.4 & 1017 \\
1,103 & -6.1 & -0.2 & 1.7 & 13.6 & 999 \\
2,075 & -2.1 & -7.1 & 7.4 & 2.3 & 1103 \\
3,679 & 1.9 & 9.3 & 7.3 & -0.84 & 1123 \\
4,033 & 2.1 & 3.6 & 5.3 & 0.86 & 1145 \\
5,257 & -2.3 & -1.7 & 3.6 & 4.3 & 1151 \\
6,811 & -2.2 & -1.2 & 2.4 & 5.1 & 1160 \\
7,279 & 1.8 & 2.9 & 5.2 & -0.3 & 1177 \\
8,631 & 1.9 & 0.9 & 3.5 & 1.4 & 1172 \\
10,737 & -1.5 & -2.2 & 4.5 & 1.5 & 1176
\end{tabular}
\caption{
Parameters $x_{1},x_{2}, x_{3}, x_{4} $ and melting temperatures $T_m$ derived from the empirical function given in Eq.~(\ref{eq:empirical}). The parameters have been determined by fitting to the global Lindemann indices obtained from MD simulation for different numbers of atoms $N$ using spherical and Wulff-shaped nanoparticles. The melting temperatures are plotted in Fig.~\ref{fig:melt_size}. 
}
\label{tab:empirical_paras}
\end{table}

\begin{figure*}[htb]
         \centering
          \includegraphics[width=0.93\textwidth]{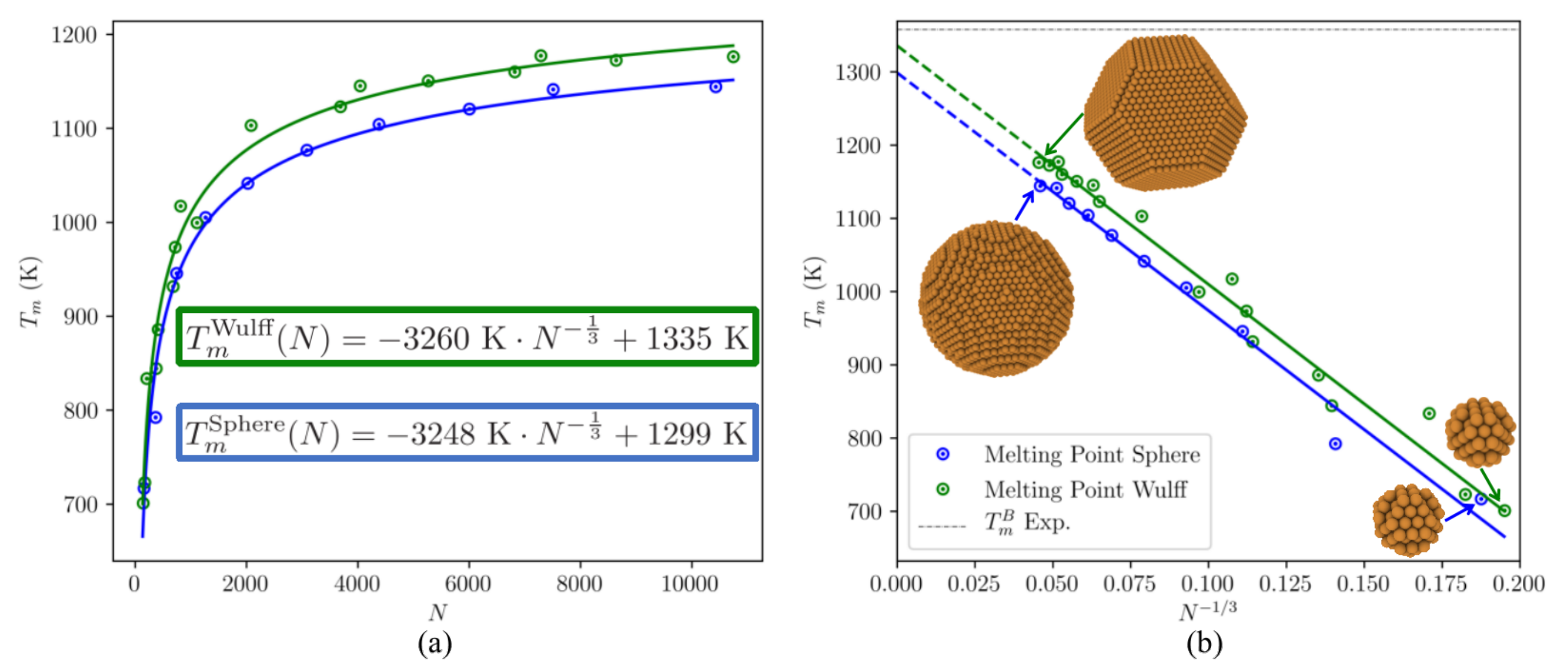}
 \caption{(a) Estimated melting temperatures $T_{m}$ of copper clusters containing different numbers of atoms $N$  determined from HDNNP-based MD trajectories using the Lindemann index (Fig.~\ref{fig:lindemannPlots}). Panel (b) shows a fit to the size scaling law Eq.~(\ref{eq:melt_law}) to extrapolate to the melting temperature $T_m^B$ of bulk copper, the corresponding equations are shown in panel (a). Two sets of clusters have been used with different initial shapes  corresponding to a Wulff construction or approximately spherical, respectively.}
    \label{fig:melt_size}
\end{figure*}

Even though Eq.~(\ref{eq:melt}) does not explicitly take the shape of the nanoparticle into account this size-scaling rule is usually still surprisingly accurate -- even in cases when the shape deviates substantially from a sphere\cite{doi:10.1063/1.2187950}.
A fit of Eq.~(\ref{eq:melt_law}) to the melting temperatures of nanoparticles with initial Wulff shapes as shown in Fig.~\ref{fig:melt_size} results in an estimated value for the bulk melting temperature of $T^{\text{Wulff,B}}_{m} = \SI[separate-uncertainty = true]{1335(13)}{\kelvin}$. 
Starting from spherical nanoparticles we find a value of $T^{\text{Sphere,B}}_{m} = \SI[separate-uncertainty = true]{1299(15)}{\kelvin}$.
The difference results from our observation that the nanoparticles with a Wulff shape generally melt at a higher temperature than the nanoparticles with a spherical shape, which can be rationalized by the wider range of atomic coordinations at the surface of the spherical particles facilitating the melting process.

Both values from the different shapes are in good agreement with the experimental bulk melting temperature of copper $T_{m}^{B} =\SI{1357.77}{\kelvin}$\cite{Haynes2014}. Still, this agreement might to some extent be the result of error compensation as the estimated bulk melting temperatures have a rather high uncertainty, because of our simple approach to melt the clusters in standard MD simulations. This is confirmed by the reported DFT value of $T_{m}^{B} =\SI[separate-uncertainty = true]{1176(100)}{\kelvin}$ obtained with the PW91 functional~\cite{meltcopper} that typically yields results very similar to the PBE functional employed in the present work. We thus assume that a lower value would be obtained also with the HDNNP by following a more rigorous simulation setup for the determination of the melting temperature of bulk copper directly.

\subsubsection{Energy and Force Prediction of Brass Clusters}

The HDNNP has to describe a wide range of local atomic environments of brass including very different copper to zinc ratios. Here we use a cluster size of $N=216$ atoms, which is still accessible by DFT calculations but large enough for complex structural features, for testing the reliability of energy and force predictions for a set of geometries not included in the reference set. This is important as the information contents of the RMSE values is often limited and allows to assess the quality of the PES only for the available DFT data, while other equally important configurations may be missed. 

The geometries used for a further validation have to cover many different structural features and compositions, which we achieve by employing simulated annealing
\cite{Kirkpatrick671} (SA) and simulations in the Semi-Grand Canonical ensemble\cite{P5442,P5443,P5715}. While the first method allows us to sample non-equilibrium geometries, the latter allows for changing both the composition of the system as well as the spatial distribution of elements, while maintaining the total number of atoms in the system constant.
We note that here we use the SGCE simulations for validation purposes only, while one of our main goals is to apply this method for a detailed investigation of the properties of brass clusters. These results will be reported elsewhere~\cite{JanPaper2}, and here we only use it as a tool to sample different concentrations of the cluster in a single simulation. 

In the SGCE the system is effectively extended by Cu and Zn particle reservoirs, and
one-to-one Metropolis Monte Carlo (MMC) exchanges of atoms between the system and the reservoirs are performed, controlled by the chemical potential difference $\Delta \mu$ between the two  particle reservoirs.
We generate random geometries of the $N=216$ cluster using the combined SA and MMC approach described above.
After some testing we found that a fixed chemical potential difference of $\Delta \mu = \SI{-2.28}{\electronvolt}$ leads to a concentration in the $\alpha$-brass regime and the value of $\Delta \mu$ was then fixed for this simulation.
The temperature of the Nos\'{e}-Hoover chain thermostat\cite{Nose1984, Hoover1996} of the molecular dynamics simulation is set to the same temperature that enters the MMC acceptance criterion. 

To generate a random initial geometry for the SA we first heat and equilibrate the cluster at $T = \SI{1000}{\kelvin}$, which is above the melting temperature of clusters of this size (see Fig.~\ref{fig:melt_size}). We then gradually cool to $\SI{200}{\kelvin}$ with
temperature-steps of $\Delta T = \SI{5}{\kelvin}$ in a series of MD simulations using the HDNNP and Nos\'{e}-Hoover chain thermostats\cite{Nose1984, Hoover1996} to control the temperature.
Each step consists of an MD run at fixed temperature and is $\SI{3}{\pico \second}$ long, with an integration time step of $\Delta t = \SI{1}{\femto \second}$, resulting in a total simulation time of about $\SI{0.5}{\nano \second}$.
From the obtained trajectory we extract a series of 150 structures covering a potential energy interval of about $\SI{0.5}{\electronvolt}$/atom and a zinc atom fraction between $25$ - $45~\%$.

\begin{figure*}[htb]
    \centering
    \includegraphics[width=0.9\textwidth]{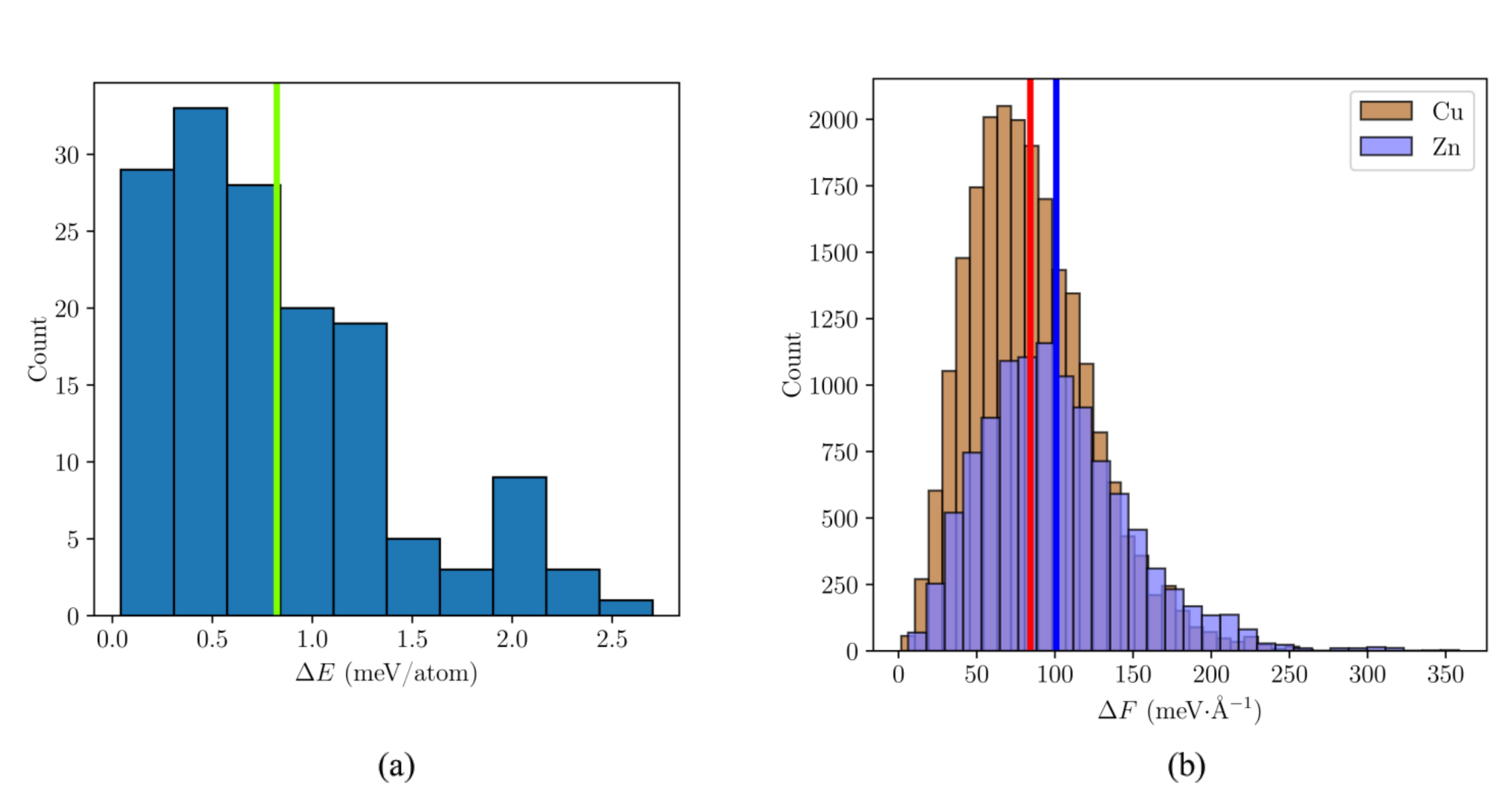}
         \caption{Distributions of the errors of the predicted cluster energies and atomic forces in comparison to DFT. Panel (a) shows a histogram of unsigned potential energy differences $\Delta E = \vert E_{\text{HDNNP}} - E_{\text{DFT}} \vert $ of 150 clusters which differ not only in their geometry but also in their chemical composition. They contain $N=216$ atoms resulting in a diameter of about $d=\SI{15}{\angstrom}$ and have been generated by simulated annealing using the HDNNP. About 80 \% of the structures have an energy error below the test set RMSE of $\SI{1.7}{\milli \electronvolt}$/atom. The green vertical line in (a) indicates the average error. 
         The distribution of $150 \times 216$ = 32,400 force errors defined in Eq.~(\ref{eq:fdev}) and their average deviations of $\SI{84.3}{\milli \electronvolt \per \angstrom}$ and $\SI{100.9}{\milli \electronvolt \per \angstrom}$ for copper and zinc atoms respectively indicated by vertical lines with blue for zinc and red for copper are shown in (b).}
    \label{fig:dftandnnpNew}
\end{figure*}

A comparison of the energies and forces predicted by the HDNNP with the corresponding values recomputed after the simulation by DFT (see Fig.~\ref{fig:dftandnnpNew}a) shows that the HDNNP predicts the PES with a very high accuracy.
The deviation between the HDNNP and DFT energies is for more than $80~\%$ of the cluster configurations below $\SI{1}{\milli \electronvolt}$/atom (see Fig.~\ref{fig:dftandnnpNew}a). 
The histogram also shows that there are only a few structures with energy errors larger than the test set RMSE of $\SI{1.7}{\milli \electronvolt}$/atom.

Next we compare the forces by computing the norm of the difference vectors between the HDNNP and DFT for every atom as follows,
\begin{align}
\Delta F =  ||\vec{F}_{\text{atom}}^{\rm HDNNP} - \vec{F}_{\text{atom}}^{\rm DFT} || ~,
 \label{eq:fdev} 
\end{align}
where $||\dots||$ is the Euclidean norm.
We find average errors of $ \langle \Delta F^\text{Copper} \rangle= \SI{84.3}{\milli \electronvolt \per \angstrom}$ and  $\langle \Delta F^\text{Zinc} \rangle  = \SI{100.9}{\milli \electronvolt \per \angstrom}$ for the force deviations of the copper and zinc atoms respectively where $\langle \dots \rangle$ the average over all atoms. The distribution of $\Delta F$ is shown in Fig.~\ref{fig:dftandnnpNew}b. 
The accuracy for the forces is about the same for all force components within the covered range between $0$ and $\SI{5}{\electronvolt \per \angstrom}$.

\subsubsection{Cohesive Energies of Brass Clusters}
\label{sec:cohe_brass}

\begin{figure}[h!]
         \centering
          \includegraphics[width=0.5\textwidth]{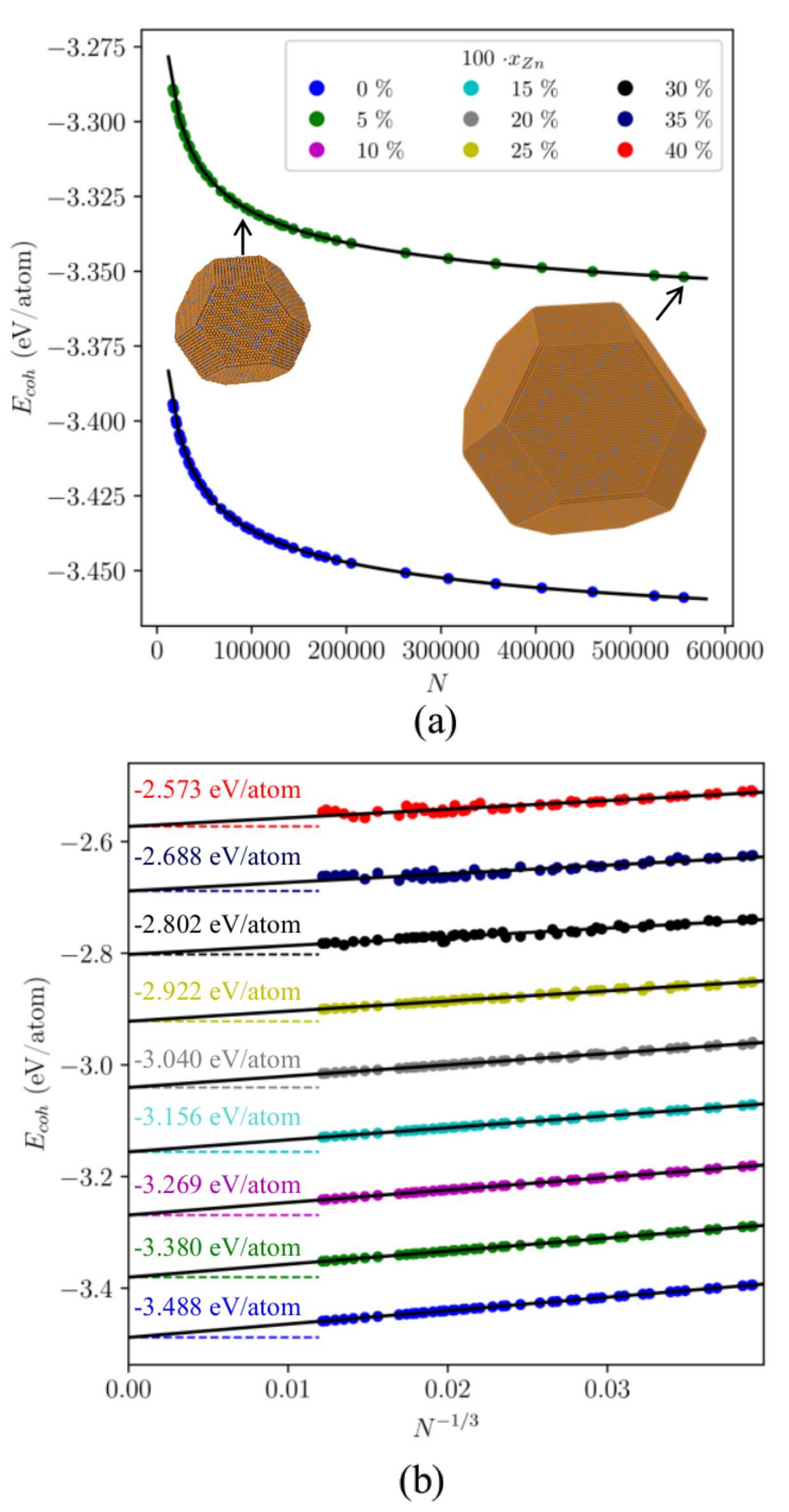}
 \caption{
 Cohesive energies $E_{\rm coh}$ of brass nanoparticles containing up to 560,000 atoms for different zinc atom fractions $x_{\text{Zn}}$. (a) shows $E_{\rm coh}$ as a function of the number of atoms $N$ for $x_{\textrm{Zn}}=0$ and $x_{\textrm{Zn}}=0.05$.
 The linear relation between $E_{\rm coh}$ and $N^{-\frac{1}{3}}$ is displayed in (b). The cohesive energies $E_{\rm coh}^{N \rightarrow \infty}$ of the infinitely large system $N \rightarrow \infty$ are the intercepts values (above each dashed line) of the linear regressions in (b). 
 }
    \label{fig:size_cohe}
\end{figure}

The cohesive energies of nanoparticles must follow the same $N^{-\frac{1}{3}}$ size scaling law as the melting temperatures in Eq.~(\ref{eq:melt_law}).
We have explicitly tested this relation using the HDNNP by performing geometry optimizations of large brass nanoparticles containing up to 560,000 atoms with an initial shape derived from a Wulff construction, which have diameters up to $\SI{24}{\nano \meter}$. As no global structural search has been performed, the initial shape is approximately conserved in these optimizations and only close local minima are found. The HDNNP cohesive energies confirm the expected scaling law (see Fig.~\ref{fig:size_cohe}).

Next, the scaling law $E_{\text{coh}} \propto N^{-\frac{1}{3}}$ has been used to estimate the bulk brass cohesive energies $E_{\text{coh}}^{\text{Bulk}}$ as a function of the zinc atom fractions $x_{\rm Zn}$. For this we use the intercept values $E_{\rm coh}^{N \rightarrow \infty}$ obtained by linear regression to the data of the cohesive energy as a function of $N^{-\frac{1}{3}}$ as shown in Fig.~\ref{fig:size_cohe}b. 
The initial geometries of the brass nanoparticles have been generated as described in the previous sec.~\ref{sec:wulff} and the zinc atoms are located at random positions of the lattice. However, the relaxation of the lattice does not include any MMC exchange moves and consequently the initial site occupations by copper and zinc are maintained. 

Finally, we compare the extrapolated cohesive energies $E_{\rm coh}^{N \rightarrow \infty}$ of large brass nanoparticles obtained from the HDNNP with the DFT results for bulk brass $E_{\text{coh}}^{\text{Bulk}}$ obtained by relaxation of $2 \times 2 \times 2$ cells containing $N=32$ atoms and the same zinc atom fractions $x_{\text{Zn}}$ as described in sec.~\ref{sec:validation}. 
As for the bulk case we find a linear increase of the cohesive energies of the brass clusters since Zn-Cu interactions in fcc geometry are energetically less favorable than Cu-Cu interactions which was also confirmed by our DFT calculations for brass bulk.
However, we find that the deviation between these values increases slightly with zinc atom fraction (see Fig.~\ref{fig:ecoh_inf}) but remain overall very small.
Further, Fig.~\ref{fig:ecoh_inf} shows that for the complete zinc atom fraction range from $0$ - $40~\%$, the HDNNP bulk cohesive energies $E_{\rm Coh}^{\rm Bulk,~HDNNP}$ obtained by relaxation of the bulk brass cell are in good agreement with the corresponding DFT values $E_{\rm Coh}^{\rm Bulk,~DFT}$.

\section{Conclusion}  \label{sec:conclusion}

In this work a DFT-based HDNNP has been constructed for the copper-zinc system which is applicable to bulk and surface structures of $\alpha$-brass and most notably also to brass nanoparticles starting from about 75 up to very large numbers of atoms. The accuracy of the HDNNP has been thoroughly validated for all these systems, and a very good agreement between the HDNNP and DFT has been found for a variety structural as well as energetic properties. While an extension of our present potential to other phases of brass is in general straightforward, a reliable description of further phases is expected to require an extension of the training set to include further atomic environments that are not relevant for the present work.

Apart from simple properties like crystal structures and surface energies, we have shown that the HDNNP can reproduce the correct size scaling behavior $T_{m} \propto E_{\text{coh}} \propto N^{-\frac{1}{3}}$ for the melting temperatures of copper nanoparticles and for the cohesive energies of brass nanoparticles.

We conclude that the constructed HDNNP is capable of providing the PES of large copper and brass nanoparticles at DFT level accuracy making it applicable to large-scale molecular dynamics and Monte Carlo simulations that can be used to compute the properties of these clusters in detail. The results of these simulations will be reported elsewhere~\cite{JanPaper2}.

\begin{acknowledgement}
We thank the Deutsche Forschungsgemeinschaft (DFG) for financial support (Be3264/10-1, project number 289217282 and INST186/1294-1 FUGG, project number 405832858). JB gratefully acknowledges a DFG Heisenberg professorship (Be3264/11-2, project number 329898176). We would also like to thank the North-German Supercomputing Alliance (HLRN) under project number nic00046 for computing time.
\end{acknowledgement}

\bibliography{literature}

\end{document}